\documentclass[12pt]{article}
\usepackage[letterpaper,margin=1in]{geometry}
\usepackage[T1]{fontenc}
\usepackage{lmodern}
\usepackage{microtype}
\usepackage{graphicx}
\usepackage{booktabs,longtable,array}
\usepackage{amsmath,amssymb}
\usepackage{seqsplit}
\usepackage{xurl}
\usepackage[authoryear,round]{natbib}
\usepackage{caption}
\usepackage{placeins}
\usepackage{setspace}
\usepackage{fancyhdr}
\usepackage{needspace}
\usepackage{etoolbox}
\pretocmd{\section}{\Needspace{9\baselineskip}}{}{}
\pretocmd{\subsection}{\Needspace{8\baselineskip}}{}{}
\usepackage[unicode,hidelinks]{hyperref}
\usepackage{bookmark}
\newif\ifreviewcopy
\reviewcopyfalse 
\ifreviewcopy
  \usepackage{lineno}
  \AtBeginDocument{\linenumbers}
\else
\fi

\newcommand{\PanelFigure}[4]{%
  \begin{figure}[!htbp]\centering
  \includegraphics[width=#4\linewidth]{figures/#1.pdf}
  \caption{#2}\label{fig:#3}
  \end{figure}}
\hypersetup{pdftitle={How much does the map of science depend on the embedding model?},
  pdfsubject={Manuscript and supplementary material},pdfauthor={Alejandro Treny Ortega}}

\hypersetup{pdfsubject={Research article},pdfauthor={Alejandro Treny Ortega}}
\begin{document}
\begin{center}
{\LARGE\bfseries How much does the map of science depend on the embedding model?\par}
\vspace{12pt}
{\large Alejandro Treny Ortega\par}
\smallskip
University College London, London, United Kingdom\par
\medskip
{\small 21 September 2026\par}
\end{center}

\begin{abstract}
Science maps describe relationships between papers. Text embedding models build these relationships by turning a paper's text into a list of numbers that gives it a position. Would another model give the same picture? I compare ten models using titles and abstracts of 500,000 publications from the OpenAlex bibliographic catalogue, published in 2000--2024 across 26 disciplines, including Medicine and Physics. Representing each discipline by its papers' average position produces similar arrangements across models, including when other papers are selected. Yet within a discipline and period, two models share, on average, only eight of the 25 papers closest to a given paper. Supplying titles alone also changes much of that list, with effects that vary between models. Models can also disagree about which discipline's papers are more spread out. One measure of spread compares angles between lines drawn from a common point to each paper's position. Moving that point can reverse which discipline appears more spread out, even when selecting other papers leaves the answer unchanged. The broad arrangement can remain similar while answers about nearby papers and differences between disciplines depend on the model, the text it receives and how its output is processed.
\end{abstract}

\section{Introduction}
A science map can help a reader find related papers or compare research areas. Its usefulness depends on the relationships it shows. When a model determines those relationships from text, choosing the model becomes part of choosing the map. Two models may lead a reader to different papers near the same starting article, or disagree about how widely an area's papers are spread. I ask how much the answer changes when we keep the papers and use another model to represent them.

The models studied here read a paper's title and abstract and turn that text into a list of numbers, called a vector or \emph{embedding}. The numbers act as coordinates: together they give the paper a position from which distances to other papers can be calculated. Papers placed close together are treated as related. This approach has been used to map biomedical literature and the structure of physics \citep{refbe130e2973,refc0e2cc29e0}. But the position comes from the model. It is not a relationship observed independently of it, and the title and abstract do not contain everything in the paper.

This concern has a history in science mapping. \citet{refa0d2d707e6} asked how much a reconstructed thematic structure reflects scientific knowledge and how much it reflects the method used to recover it. \citet{ref6a2f5912bc} found that methods could obtain similar overall accuracy while assigning different papers to clusters. An overall score can therefore leave an important question unanswered: would another method support the particular comparison a reader wants to make?

I use ten models and a fixed collection of 500,000 publications to examine three questions. The first concerns shared relationships: do models preserve the broad arrangement of areas and identify the same nearby papers? The second asks how much of a change comes from what the model reads: does supplying only a title matter as much as switching models? The third, added as a later exploration, concerns a description we might draw from the map: do models agree that one area's papers are more spread out than another's? Repeating each type of comparison with other selected articles also shows whether its answer depends on the particular papers used.

The contribution is to follow the comparisons that a reader might draw from a science map, and to test which choices they survive. A claim that one area's papers are more spread out than another's can repeat when different papers are selected, yet receive opposite answers from different models. I examine these persistent disagreements alongside the size of the differences and their dependence on processing choices. This connects a broad assessment of model agreement to the reliability of particular comparisons between research areas.

\section{Background and research questions}
\subsection{What earlier comparisons tell us}
Previous studies assess scientific embeddings against several useful criteria. \citet{refc0e2cc29e0} compare them with disciplinary classifications in physics; \citet{refbe130e2973} examine related biomedical papers and their labels. Such checks ask whether a representation agrees with information about the literature. The present study asks a complementary question: which relationships remain when the representation changes?

The model is only one choice involved. The \emph{input text} is the text we give it to read: a title alone conveys less information than a title followed by its abstract. Models also accept different lengths of text and use different rules to produce a single vector for it. \citet{singh-singh-2022-inefficiency} show that scholarly search can be sensitive to the wording of short queries. The supplied text therefore belongs in the comparison alongside the model.

Several parts of this comparison have direct precedents. \citet{ref4e1e0d5582} combine structural similarity with overlap in retrieved documents. \citet{ref6da178065c} examine local geometry in scientific literature, while \citet{ref483cb38608} distinguishes representational similarity from the stability of geometry. Here, I bring these concerns to specific comparisons between research areas. For each pair of areas, I ask whether a model keeps the same ordering across article selections, whether another model persistently supports the opposite ordering, and whether those same models retain their answers after a change in processing. Matched text and search comparisons then help establish which other choices affect the result.

\subsection{What agreement can establish}
Two changes need to be kept separate. To measure \emph{agreement}, I keep the articles fixed and compare models. To check \emph{selection stability}, I keep the model fixed and choose other articles under the same rules. An \emph{article selection} is simply a set chosen from the existing collection; repeating it does not mean downloading new papers. A result can repeat with other articles in each model while the models consistently give opposite answers.

Neither property establishes that the answer is correct about the research itself. Models can share an error, and a repeatable pattern can be hard to interpret. In a study of invasion biology, \citet{QSSreviewQ17} show why topic solutions require substantive interpretation beyond numerical checks. Here, agreement is evidence that a result survives a change of model, not evidence of thematic truth.

The word map needs one further distinction here. I compare the numerical positions, each with hundreds of coordinates, before reducing them to a drawing. A two-dimensional plot would add a transformation that could change apparent distances. \emph{Structure} means the pattern of relationships among the objects being compared, whether papers or area summaries. \emph{Spread} describes how far apart the representations lie within one area. Two models can share much of the broad structure without agreeing on every area's spread.

\Needspace{16\baselineskip}
\subsection{Research questions}
\textbf{RQ1.} When the model changes, what remains of the arrangement of research areas and of the papers placed close together?

\textbf{RQ2.} How much does changing the input text, or the rule used to form one paper vector, matter compared with changing the model?

\textbf{RQ3.} Do models agree about which area's paper representations are more spread out, or whether that spread is concentrated along a few directions rather than many? Do these comparisons repeat with other articles?

The third question was added after the initial comparisons and is exploratory. All three concern a fixed collection and a fixed set of models. They do not seek a winning model or estimate the structure of all scientific literature.

\section{Data and methods}
\subsection{One collection, several questions}
OpenAlex is a bibliographic catalogue that supplies titles, abstracts and other information about publications \citep{priem2022openalex}. The starting collection, or corpus, contains 500,000 distinct work IDs retrieved and cleaned by 15 September 2026. Records were labelled as English articles, reviews or conference papers published in 2000--2024. Each required a title, an abstract of at least 50 words reconstructed from OpenAlex's stored format, and a primary research area. Retractions, ancillary material and clear content failures were excluded. Less certain cases and possible duplicate versions were retained with review flags. Those flags record a concern for later checks; they do not establish that every retained record is correct. Supplement S1 gives the rules and counts.

OpenAlex assigns papers to broad \emph{Fields} and narrower \emph{Subfields}, using information from text, citations and journals \citep{openalex2026topics}. I call these areas and specialties. For example, Medicine is an area; Cardiology and Cardiovascular Medicine is one of its specialties. The 26 areas provide the main units of comparison. These database labels organise the analysis; they are not independently verified descriptions of every paper.

The collection has two parts. Reproducible random requests to OpenAlex supplied 400,000 eligible records. Another 100,000 were added to strengthen groups that initially had few papers from the same area and five-year period. For example, Medicine in 2000--04 is one such group. The five periods are 2000--04, 2005--09, 2010--14, 2015--19 and 2020--24. Combined with the 26 areas, they give 130 area--period groups, each with at least 2,165 papers. Searches across the whole general collection use the original 400,000 records; comparisons within areas can use both parts. The reinforcement makes smaller groups easier to compare, but changes their share of the collection. Neither that share nor an average giving every area equal weight estimates world publication proportions.

All ten models represent the 500,000 papers using titles and abstracts, but each subsequent question uses the articles it needs. The \emph{input-text experiment} compares titles, abstracts and both together on the same 52,000 papers: 2,000 per area, with 400 from each period. These include the 26,000 used in an initial trial. The expansion tests how the agreement measures and their sensitivity to article selection change as more papers are included. The later study of spread within areas reuses this set and its title-plus-abstract vectors. The \emph{common-passage check} uses a different set of 52,000 papers to ensure that every model reads exactly the same passage, short enough to fit all input limits. This separates a choice about content from a choice about length. Table~\ref{tab:T01} is a guide to these comparisons; its rows overlap and should not be added together. Within any comparison, the models always represent the same paper IDs.
\begin{table}[!htbp]
\par\noindent\begin{minipage}{\linewidth}
\small
\setlength{\tabcolsep}{4pt}
\renewcommand{\arraystretch}{1.10}
\captionof{table}{Questions, comparisons and article sets.}\label{tab:T01}
\begin{tabular}{@{}>{\raggedright\arraybackslash}p{4.0cm}>{\raggedright\arraybackslash}p{5.4cm}>{\raggedright\arraybackslash}p{5.75cm}@{}}
\toprule
\textbf{Question} & \textbf{What is compared} & \textbf{Articles used} \\
\midrule
Does the broad arrangement stay similar? & Ten models: relationships between groups, summarised by their average positions, and between papers within groups. & 256 papers per group to calculate its centre; repeated choices and size checks at 128 and 512. \\
\addlinespace[8pt]
Do models find the same nearby papers? & How many of the 25 closest papers appear in both models' lists. & All papers within each area and period; a separate check fixes 2,048 candidates per group. Global search: 13,000 starting papers among 400,000 candidates. \\
\addlinespace[8pt]
Does changing the input text matter? & The same model reads only the title, only the abstract, or both. & 52,000: 2,000 per area, 400 per period. Includes an initial 26,000-paper trial. \\
\addlinespace[8pt]
Does reading the same passage matter? & Each model's usual input versus a common passage that fits every model. & A different set of 52,000, with 400 per area and period. \\
\addlinespace[8pt]
Do models agree on how papers spread within areas? & Angles between paper vectors and how their variation spreads across directions; all 325 area pairs under ten models. & The same 52,000 as in the input-text experiment; further article choices test whether the comparisons repeat. \\
\addlinespace[8pt]
Does searching a narrower specialty matter? & For example, Cardiology and Cardiovascular Medicine versus Medicine, with 256 candidates in each search. & 50 fixed starting papers per specialty; ten choices of the other 206 candidates. In total, 10,850 starting papers across 217 specialties. \\
\bottomrule
\end{tabular}
\par\smallskip
All sets come from the same 500,000-paper collection: 400,000 general records plus 100,000 added to strengthen small area--period groups. Rows are not added together. Models use the same paper IDs within each comparison. Candidates are the papers available for a neighbour search. In the scope comparison, both candidate sets include the same 50 starting papers. Methods and Supplements S1--S4 and S9 give the full controls. The 1,300 software-test papers are not another scientific sample.
\end{minipage}\par
\end{table}

\subsection{How each model represents a paper}
The ten models in Table~\ref{tab:T02} were already trained before this study. I use their published versions to convert text into vectors, without training them again on this collection. They cover scientific documents, general sentence similarity, and scientific or biomedical language. I prioritised models used most often in the science-mapping studies reviewed, such as SPECTER and SciBERT. I added models used as comparators in those studies to cover different training approaches \citep{refbe130e2973,refc0e2cc29e0,ref07e33b3339}. This is a fixed comparison set, not a random sample of models or a ranking of the newest systems. Versions, input formats and software settings were fixed before calculating the vectors.

The main input is the title followed by the abstract, in each model's usual format. Models divide text into small pieces called \emph{tokens}, which can be whole words or parts of words. A token limit is therefore not a word limit. Under their standard settings, MiniLM reads at most 256 tokens and MPNet 384; text beyond that limit is cut off. This affects 49.58\% and 20.16\% of the papers, respectively. I retain those settings for the main comparison. Because MiniLM cuts off text in almost half the papers, I also test it separately at 512 tokens to check whether reading more text changes its agreement with the other models. The common-passage check addresses the resulting difference in what models can read: all receive the same content, although each splits it into tokens in its own way.

A further choice concerns how a model produces one vector for the whole paper. BERT, SciBERT, BioBERT and PubMedBERT produce an output vector at each token position, but were not trained specifically to compare whole documents. The main rule averages those outputs after the model has finished processing the text. This step is called \emph{mean pooling}. The average includes special symbols used to mark the text and excludes empty positions added as padding. Two alternatives take only the output at the first special-symbol position (CLS) or at the final non-padding position (SEP). These outputs have processed the surrounding text; they do not represent only the first or last word. The mean rule was specified before the primary comparisons and retained after early checks, rather than chosen to maximise agreement. The other six models use their documented rules. Table~\ref{tab:T02} and Supplement S1 specify the exact settings.
\begin{table}[!htbp]
\par\noindent\begin{minipage}{\linewidth}
\small
\setlength{\tabcolsep}{4pt}
\renewcommand{\arraystretch}{1.13}
\captionof{table}{Embedding models and primary representation settings.}\label{tab:T02}
\begin{tabular}{@{}>{\raggedright\arraybackslash}m{3.3cm}>{\centering\arraybackslash}m{1.6cm}>{\centering\arraybackslash}m{1.65cm}>{\raggedright\arraybackslash}m{3.0cm}>{\raggedright\arraybackslash}m{4.9cm}@{}}
\toprule
\textbf{Model} & \textbf{Vector size} & \textbf{Token limit} & \textbf{Paper vector} & \textbf{Reference} \\
\midrule
\textbf{SPECTER} & 768 & 512 & CLS & \citep{cohan-etal-2020-specter} \\
\addlinespace[2pt]
\textbf{SPECTER2} & 768 & 512 & CLS + adapter & \citep{singh-etal-2023-scirepeval} \\
\addlinespace[2pt]
\textbf{SciNCL} & 768 & 512 & CLS & \citep{ostendorff-etal-2022-neighborhood} \\
\addlinespace[2pt]
\textbf{SciBERT} & 768 & 512 & Mean & \citep{beltagy-etal-2019-scibert} \\
\addlinespace[2pt]
\textbf{BERT} & 768 & 512 & Mean & \citep{devlin-etal-2019-bert} \\
\addlinespace[2pt]
\textbf{MPNet} & 768 & 384 & Mean & \citep{reimers-gurevych-2019-sentence} \\
\addlinespace[2pt]
\textbf{MiniLM} & 384 & 256 & Mean & \citep{reimers-gurevych-2019-sentence} \\
\addlinespace[2pt]
\textbf{PubMedBERT / BiomedBERT} & 768 & 512 & Mean & \citep{reff9ebfb9307} \\
\addlinespace[2pt]
\textbf{BioBERT} & 768 & 512 & Mean & \citep{ref4f3e28932e} \\
\addlinespace[2pt]
\textbf{SimCSE} & 768 & 512 & CLS & \citep{gao-etal-2021-simcse} \\
\bottomrule
\end{tabular}
\par\smallskip{\small Vector size is the number of coordinates, not the number of words read. MPNet and MiniLM use all-mpnet-base-v2 and all-MiniLM-L6-v2, respectively. Limits count tokens, the pieces into which a model divides text, including its special symbols. Mean averages the final token outputs (the last layer), including special symbols and excluding empty padding positions. CLS uses the first-position output before the additional transformation called the pooler. SPECTER2 also uses its proximity adapter, an additional trained component. For MPNet/MiniLM, the reference describes the Sentence-BERT framework, not the exact training of these published versions. Changing the pooling rule does not add a separately trained model. Full revisions, formats and truncation: Table S2.\par}
\end{minipage}\par
\end{table}

\subsection{Following the map from areas to individual papers}
A vector can be pictured as an arrow from a common starting point to the paper's position. Before comparison, I scale every arrow to length one, keeping the direction in which it points. This operation is called \emph{normalisation}; it prevents differences in vector length from determining closeness. To summarise an area, I average its article vectors and also scale the resulting vector to length one. I call this summary its \emph{centre}. Comparing the centres asks whether models place areas in similar relationships to one another. Comparing the individual papers inside an area asks how much of its internal organisation they share.

The first centre comparison fixes 256 papers per group, so that unequal group sizes do not drive the comparison. It considers the 26 areas, the 217 specialties, and a restricted set containing one specialty from each area. To test dependence on which papers were chosen, I repeat article selection fifty times while keeping each group's counts from the five publication periods. To test dependence on size, I use 128, 256 and 512 papers for the same 26 areas and the 183 specialties that have enough papers at every size. These are comparison settings, not a claim that one size is universally sufficient.

Each selection also supplies a reference: the same papers are randomly redistributed into groups while preserving group sizes and publication-period counts. The reference asks how much model agreement remains when groups no longer follow the area or specialty labels. Further checks leave out one area at a time and, for the restricted specialty set, separate choosing a specialty from choosing its papers. Supplement S9 details these centre checks.

\subsection{Finding and comparing nearby papers}
For an individual paper, its \emph{neighbours} are the papers placed closest to it. The \emph{starting paper} is the one whose neighbours we seek; the \emph{candidates} are all papers available for that search. Each model returns the closest 25, and agreement is the fraction that appears in both models' lists. Lists of 10 and 50 provide checks. Closeness uses cosine similarity: arrows pointing in more similar directions count as closer. The starting paper cannot be its own neighbour. The lists are calculated exactly and checked by independent sorting.

A search depends on the papers available to it. Searching among every paper from the same area and period gives different groups different numbers of candidates. A separate check fixes 2,048 candidates per group to remove that difference in count. Within a group, every model searches the same papers. A further search uses the 400,000 general records without area or period restrictions. These comparisons answer different questions, so each result specifies its candidate set.

The quality check asks whether flagged records and duplicates explain disagreement between models. Applying the existing review flags and duplicate rule leaves 448,886 records. I compare neighbour lists for the same retained starting papers before and after filtering the candidates. Removing records also reduces search size, so further checks hold that size fixed: 1,024 candidates in all 130 groups, and 2,048 in the 63 groups still large enough after filtering. The before-and-after sets include the same 100 clean starting papers per group. Here, clean means passing the recorded filter; it does not mean that every possible source error has been ruled out.

To separate search scope from search size, I compare a specialty with its parent area using the same candidate count and matched dates. Consider 50 fixed starting papers from Cardiology and Cardiovascular Medicine. For each, we search among 256 cardiology papers and among 256 papers from Medicine. Both sets include those same 50 starting papers, leaving 206 places for others; the current starting paper is excluded when its neighbours are ranked. I choose the other 206 ten times, keeping the 50 fixed and matching publication-period counts between searches. The procedure covers 217 specialties, giving 10,850 starting papers. It deliberately includes the 50 specialty papers in the broader set too. Its result therefore applies to this matched comparison, rather than to an unrestricted random selection from Medicine.

\subsection{Measuring shared structure}
To measure shared structure, I compare the pattern of relationships among the same objects under two models. The objects can be individual papers or the centres that summarise areas. The main measure is corrected linear centred kernel alignment, abbreviated CKA \citep{pmlr-v97-kornblith19a,refa961d84a4e}. It compares these patterns without requiring a coordinate in one model to mean the same thing as a coordinate in another. A value near one, such as 0.9, indicates that the patterns are very similar on this measure. It does not mean that a corresponding percentage of science is correctly represented. The correction reduces a known upward distortion associated with finite samples; it does not remove all statistical bias.

Two further measures check other aspects of shared structure. Procrustes alignment allows the representations to rotate and change scale before comparing them. The distance-rank check asks whether pairs of papers have a similar ordering from nearer to further apart. These measures need not give identical answers \citep{NEURIPS2021_252a3dba,pmlr-v285-williams24a}. Their formulas appear in Supplement S2.

\subsection{Describing spread within an area}
A different question concerns the shape formed by the papers within each area. I describe two properties. \emph{Angular spread} summarises the angles between their vectors, using the median-distance convention in Supplement S6. Picture arrows drawn from a common point to the papers: larger angles mean that the arrows point further apart.

\emph{Effective linear dimension} describes how that cloud of paper positions spreads across directions. Here, variation means differences between the papers' coordinates. A cloud concentrated along a line varies mainly in one direction; one spread across a plane varies in more. The participation ratio, or PR, extends this idea to the full vector space and allows values between whole numbers. A larger PR means that variation is less concentrated in a few directions. Neither PR nor angular spread counts research topics or establishes thematic diversity.

Alternative angle summaries and dimension measures check how much the answer depends on the formula. For dimension, the entropy-based alternative measures how evenly variation is distributed. D80 counts the fewest directions needed to account for 80\% of that variation. Because these alternatives describe the same underlying variation, agreement among them would not independently validate a thematic interpretation. I also tested a further idea: whether weak links between nearby papers could reveal an area split into separate groups. Artificial examples exposed problems with reading that signal as thematic fragmentation. Supplement S6 gives the definitions and controls.

\subsection{Checking whether a comparison repeats}
With 26 areas, there are 325 ways to choose two different areas, counting each pair once. For each property, each model gives a value for area A and a value for area B. I ask whether the same area keeps the larger value when different papers are chosen. The main calculation uses 2,000 papers per area. I repeat it with twenty half samples of 1,000 and five further selections of 2,000. An ordering is \emph{persistent} only if it holds in all 26 conditions. This refers to the order of the two measured values, not to the direction in which a vector points. The dimension alternatives use the same conditions.

Another check changes the common point from which the arrows are drawn. For each model, I calculate the mean of the balanced 52,000 unit vectors used in the input-text experiment, subtract it from every vector, and scale each resulting vector back to length one. This is \emph{global centring followed by normalisation}. The new origin is the average position in this set. I then recalculate spread and dimension to ask whether comparisons between areas depend on that choice. Supplement S6 distinguishes this change to the vectors from the internal centring already used in the PR formula.

The results fall into three classes. In \emph{unanimous agreement}, all ten models persistently give the same area the larger value. In \emph{opposition}, at least two fixed models persistently give opposite orderings. Otherwise the comparison is \emph{unresolved}. For example, some models may change their ordering with other articles and no fixed pair may maintain opposite answers. Unresolved does not mean equal. Nor does a tie in one model necessarily leave the whole comparison unresolved: two other models can still maintain opposite orderings.

I also ask whether the differences are large enough to matter numerically. Besides any nonzero difference, I require minimum differences of 1, 5 and 10\% of the two areas' average value. A model's ordering counts at a given cutoff only if its difference exceeds that cutoff in every selection. The three classification rules are then applied again. These percentages describe the size of a measured difference, distinct from the percentage of the 325 pairs assigned to a class. They are sensitivity checks, not universal thresholds of scientific importance.

A separate diagnostic, the \emph{selection alert}, marks a result that changes more than a recorded tolerance allows when we choose other articles or use another tested size. It identifies a result worth examining for sensitivity to article choice; it is neither a source-record error nor one of the three area-pair classes above. Each check has its own rule and count, retained in the supplement.

Selections overlap because they come from the same fixed collection. Their ranges describe changes among the available papers, rather than confidence intervals for all scientific literature. A central 95\% selection range leaves out the lowest 2.5\% and highest 2.5\% of the recorded results. Comparisons also reuse models and areas, so I describe their patterns without treating them as independent observations in significance tests.

The work developed in stages. The initial structure and neighbour comparisons came before the expanded text experiment and the specialty searches. The geometric analysis was a later exploration, and the summary of 325 area pairs followed its first examples. Rules were recorded before the corresponding new calculations, with earlier results already known; some controls were added along the way. Supplement S2 documents this sequence. The full study was not registered in advance with an external service, so these later checks should not be read as predictions made before any results were seen.

\Needspace{14\baselineskip}
\section{Results}
\FloatBarrier
\subsection{The broad arrangement is shared more than the detail}
Models share a similar arrangement of area centres, and choosing other papers changes that conclusion little. Each mean combines the 45 ways of pairing the ten models. For the original article selection, corrected CKA averages 0.939 for area centres and 0.911 for specialty centres. Across fifty selections, the averages are 0.940 and 0.911, with central 95\% selection ranges of 0.937--0.943 and 0.910--0.912 (Figure~\ref{fig:scales}A). Randomly redistributed groups average 0.627 and 0.636. None of their selection means reaches the lowest observed selection mean. Thus, the labelled groups produce more shared structure than these random references.

The arrangement also changes little when centres use more papers. At 128, 256 and 512 papers, area-centre means are 0.938, 0.940 and 0.941; the same 183 eligible specialties give 0.913, 0.914 and 0.915. Leaving out each area in turn gives 0.933--0.946 in the original comparison, so no single area accounts for the high agreement. When only one specialty represents each area, the mean is 0.904. In this restricted comparison, changing which specialty represents an area matters more than changing its selected papers (Table S18).
\PanelFigure{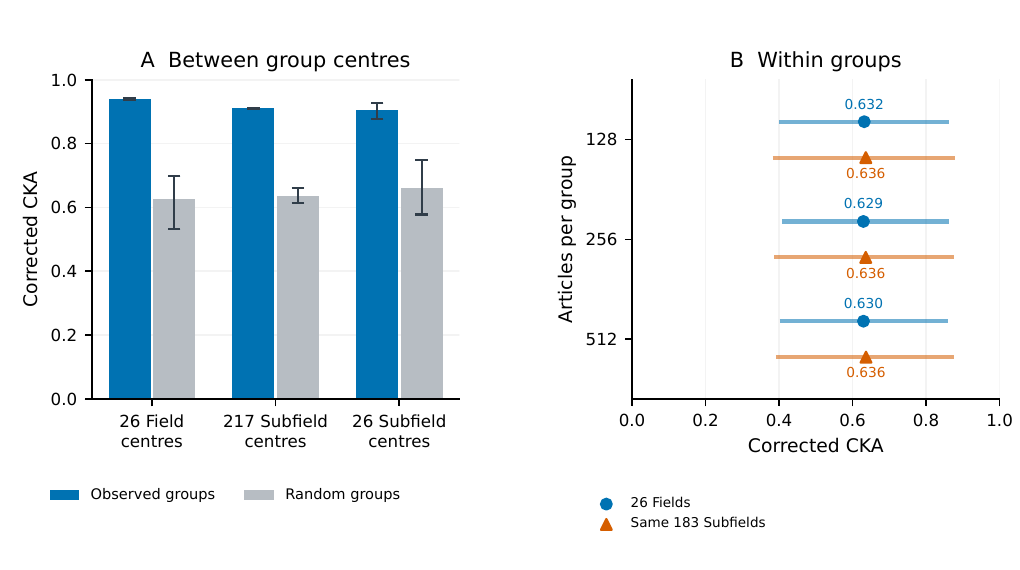}{Agreement about group centres and individual papers. Higher corrected CKA means greater agreement between models. (A) Each centre summarises 256 papers. Area and specialty results average fifty article selections. The 26-specialty condition chooses one specialty from each of the 26 areas, repeats that choice fifty times, and chooses ten article sets within each choice. Random groups redistribute the same papers, preserving sizes and dates. Lines span the central 95\% of selection means. (B) Agreement among individual papers within 26 areas and the same 183 specialties, at three article counts. Points show means; lines span the central 95\% of group--model-pair results: 1,170 for areas and 8,235 for specialties. Panel A compares relationships between group summaries; panel B compares relationships between papers inside groups. Their ranges describe different sources of variation and are not confidence intervals.}{scales}{1}

Agreement is lower when we compare the individual papers inside a group. Mean CKA stays around 0.63 within areas and 0.64 within the same 183 specialties as the count rises from 128 to 512 (Figure~\ref{fig:scales}B). That stable average hides individual comparisons that change more. Of the 8,235 combinations of a specialty and a model pair, 2,628 trigger the selection alert (31.9\%): their agreement changes more than the recorded tolerance allows across article choices or tested sizes. A stable mean therefore does not make every specialty result stable. The contrast with centres also concerns different objects---group summaries versus individual papers---so it does not isolate the effect of scale alone.

The alternative structure measures broadly agree about which model pairs are more similar. Within each of the original 130 area--period groups, I order model pairs from less to more agreement and compare those orderings. Their mean rank correlations with corrected CKA are 0.891 for Procrustes and 0.881 for distance ranks; values nearer one mean more similar orderings. Agreement is not complete. The original article-selection check also retains 50 alerts among 5,850 comparisons, corresponding to 45 model pairs in each of the 130 groups (Table S4). This is a different check from the specialty alerts above, with its own sizes and tolerances.

\FloatBarrier
\subsection{Changing the model changes which papers appear nearby}
When every paper in the same area and period is available as a candidate, two models share an average of 30.3\% of the 25 nearest neighbours: about eight papers out of 25. The average gives equal weight to each area--period group and model pair. Holding every group's candidate count at 2,048 gives 31.9\% agreement. The count is shared across groups; the actual papers are shared across models within each group, not across disciplines.

The general-collection search asks a broader question. It takes 100 starting papers from each of the 130 groups, giving 13,000 in total, and searches among the 400,000 general records. Agreement is 17.5\%. This search both admits other disciplines and periods and offers many more candidates. Because both change together, the comparison cannot tell us how much of the lower agreement comes from each change.

Filtering source flags and duplicates changes little of this picture. To make the before-and-after comparison fair, both averages use only the starting papers that survive the filter. Their initial agreement is 30.58\%, slightly different from the 30.3\% for all starting papers above, and rises to 31.43\% after filtering candidates. The gain is 0.85 percentage points. Both averages still mean roughly eight shared papers out of 25. Keeping candidate counts fixed reduces the gain to 0.28 points with 1,024 candidates and 0.19 with 2,048. The issues removed by this filter explain little of the overall disagreement, although some areas change more than others (Table S19).

The matched specialty searches isolate a change in disciplinary scope more closely: both searches have 256 candidates, the same starting papers and matched dates (Figure~\ref{fig:paired}). Searching within the specialty makes models share fewer of their 25 neighbours in 127 of 217 specialties (58.5\%), but more in the other 90 (41.5\%). The average change from the parent-area search is $-1.35$ percentage points. With lists of 50 neighbours, the split is almost even: 109 specialties decrease and 108 increase. Narrowing the search therefore does not have the same effect everywhere. These results describe how much the models' lists overlap; they do not assess how useful the listed papers are.
\PanelFigure{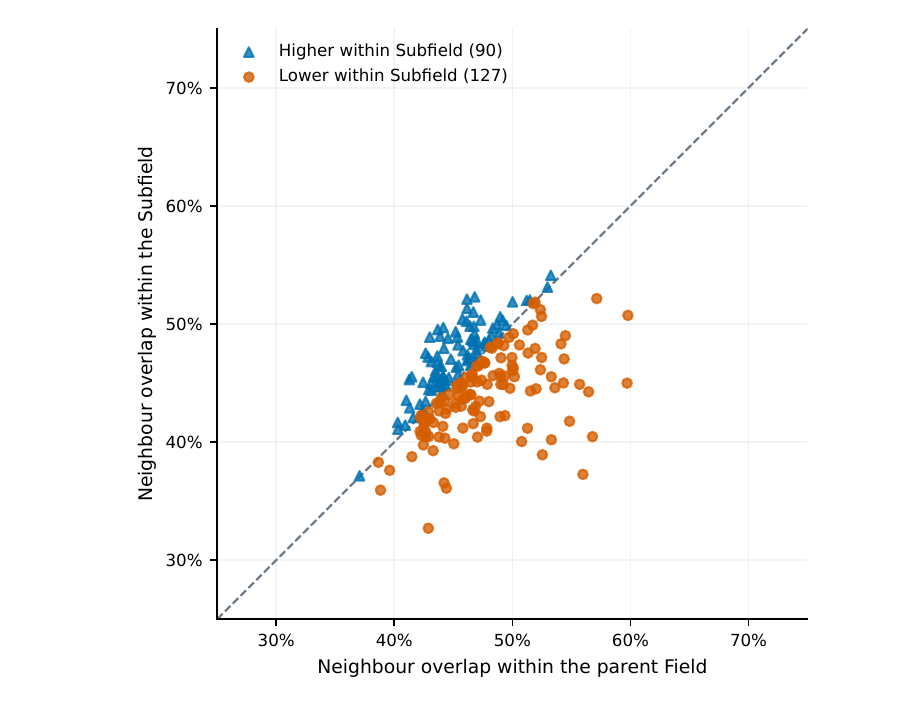}{Does searching within a specialty change agreement? Each point represents one of 217 specialties. The horizontal axis shows the fraction of neighbours shared by models when searching the parent area, such as Medicine. The vertical axis shows that fraction within its specialty, such as Cardiology and Cardiovascular Medicine. Above the diagonal, models share more neighbours in the narrower search; below it, they share fewer. Both candidate sets contain 256 papers, including the same 50 starting papers, with matched dates. Values average model pairs and ten choices of the other 206 candidates. The parent-area set always includes those 50 specialty papers. Differences between points describe specialties, not confidence intervals.}{paired}{0.90}

Pairs of models that agree more on overall structure also tend to share more neighbours. Ordering the 45 model pairs by each type of agreement gives a rank correlation of 0.920. This association does not make the two checks interchangeable: a pair can preserve broad relationships while replacing many individual neighbours. The model pairs also share models with one another, so the association is descriptive rather than evidence from 45 independent observations.

Supplement S4 moves from averages to named papers: it checks whether a paper keeps the same neighbours when sample size, the rule for forming its vector, or neighbour-list length changes. Reading these examples also revealed source problems. A history book appears under Physics, a text about bacterial infection under Politics, and another record is an announcement. I checked these against the downloaded OpenAlex records: the classifications and content were already present there, rather than introduced by matching the wrong papers across models. The examples remain in the analysis. Because they were selected as cases to inspect, their number does not estimate the frequency of such problems in the collection.

\FloatBarrier
\subsection{Similar average changes hide different model behaviour}
The input-text experiment now changes what the model reads. It uses the same 52,000 papers for three inputs: title alone, abstract alone, and title plus abstract. Here, each area contributes 2,000 candidates, combining its five periods. This differs from the preceding searches within a single area and period.

Starting from title plus abstract, I compare keeping the text and switching models with keeping the model and supplying only the title. The first replaces an average of 68.8\% of the 25 neighbours; the second replaces 70.7\% (Figure~\ref{fig:input}). These are fractions replaced, so a larger value means less agreement. The corresponding structural changes are 0.370 and 0.366, measured as one minus CKA: subtracting agreement from one makes larger values mean more change here too. The two average changes are similar, but they need not be similar for every model.
\PanelFigure{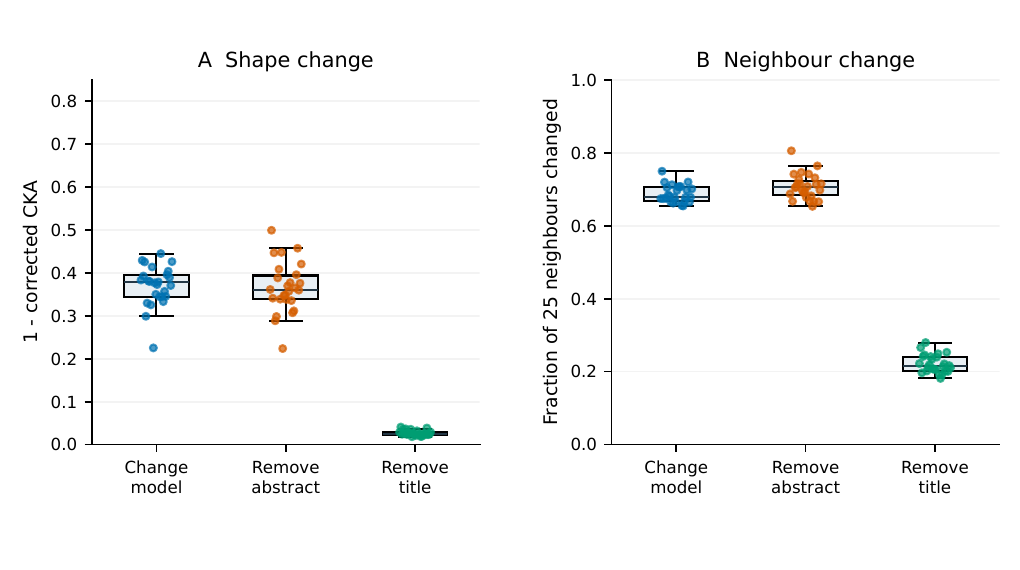}{Changing the model or changing its input. All comparisons use the same 52,000 papers, with 2,000 per area across five periods. The starting input is title plus abstract. Model change compares models on that input; text change keeps the model and uses only the title or only the abstract. Larger values mean more change: (A) one minus corrected CKA; (B) the fraction of 25 neighbours replaced among the area's 2,000 candidates. Each point is an area average. The line inside each box is the median; the box covers the middle half of area values. Whiskers extend to values within 1.5 box heights of the box edges. These distributions describe differences between areas, not confidence intervals. Models retain their usual text limits.}{input}{1}

Keeping the abstract and removing only the title changes less: 0.027 for structure and 22.1\% of neighbours. Titles alone also affect models differently, replacing 51.7\% of neighbours for MPNet and 84.2\% for SciBERT (Table S5). The similar overall averages therefore should not be read as a general equivalence between changing text and changing models.

To see whether that average comparison repeats, I choose fifty balanced sets of 1,000 candidates per area, half the full experiment's count. These smaller searches replace 64.93\% of the 25 neighbours when switching models and 66.49\% with titles alone. Title-only input therefore replaces 1.56 percentage points more, with a central 95\% selection range of 1.44--1.69. The overall difference stays positive in every repetition. It averages 1.97 points with 10 neighbours and 1.04 with 50.

The model-specific results tell a less uniform story. Differences range from $-14.04$ points for MPNet to $+16.59$ for SciBERT. A negative value means that titles alone change fewer neighbours than switching models; a positive value means they change more. Thus, a repeatable positive grand mean combines models for which the two changes have opposite relative effects (Supplement S9).

Changing the rule that forms a paper vector can also reverse the overall result. For the four word-level models---BERT, SciBERT, BioBERT and PubMedBERT---I replace the mean of token outputs with the CLS or SEP output defined above. The other six models keep their usual rules. Both sides of the comparison are recalculated: the effect of switching models and the effect of supplying titles alone. With 25 neighbours, titles alone now replace 3.13 percentage points fewer for CLS and 2.00 fewer for SEP, in every repetition. The ordering therefore depends on this processing choice. These are comparisons of practical ways to use the models; they do not isolate the causal contribution of text, model design or a single processing step.

The expansion from the first 26,000 papers to 52,000 also reduced sensitivity to article selection in the structural comparison. Applying the same check with 100 repeated selections at each size lowers alerts from 72 of 520 model--area--input contrasts (13.8\%) to five of 520 (1.0\%). Fewer contrasts therefore exceed the recorded tolerance. This does not establish stability for every neighbour comparison. Supplement S3 retains the earlier twenty-selection check separately. The common-passage and MiniLM 512-token checks instead concern input length; they do not reveal a coding error in MiniLM's original limit.

\FloatBarrier
\subsection{A comparison between two areas can reverse under another model}
The final analysis asks whether models support the same description of an area. It reuses the input-text experiment's 52,000 papers and their title-plus-abstract vectors. For example, do the arrows representing one area's papers point further apart than those of another? Before comparing areas, I check whether each area's measured value changes greatly when other papers are chosen.

Ten models and 26 areas give 260 cases to check. Angular spread has no selection alerts in either the half-sample or equal-size check. PR has four alerts (1.5\%) with twenty half samples of 1,000 papers per area, and none with five further selections of 2,000. These checks use different sizes and repetition counts. Passing the second does not remove the sensitivity detected by the first.

Repeatable values within a model do not ensure agreement between models. Among the 325 area pairs, 80.6\% have persistent opposite answers from at least two models for angular spread, and 68.0\% for PR (Figure~\ref{fig:pairs}). For example, one fixed model gives area A a larger angular spread than B in every article selection, while another gives B the larger spread in every selection. The other models need not all disagree with one another. ``Larger'' refers to the measured property, not to a higher position in a drawing or better research.

Unanimous agreement is less common: all ten models persistently give the same area the larger value in 12.9\% of pairs for angular spread and 16.6\% for PR. The remaining 6.5\% and 15.4\% do not meet either the unanimous-agreement or opposition rule. They remain unresolved under those rules.

Many reversals concern small differences between areas. Requiring each persistent ordering to exceed a 5\% difference, relative to the two measured values' average, reduces opposition to 29.5\% of the 325 pairs for angular spread and 54.5\% for PR. At a 10\% minimum difference, the shares fall to 5.8\% and 38.8\%. The first percentage in each comparison is a requirement on the difference's size; the resulting percentage tells us how many pairs still meet it. A reversal over a tiny difference and one over a larger difference should not carry the same interpretation.
\PanelFigure{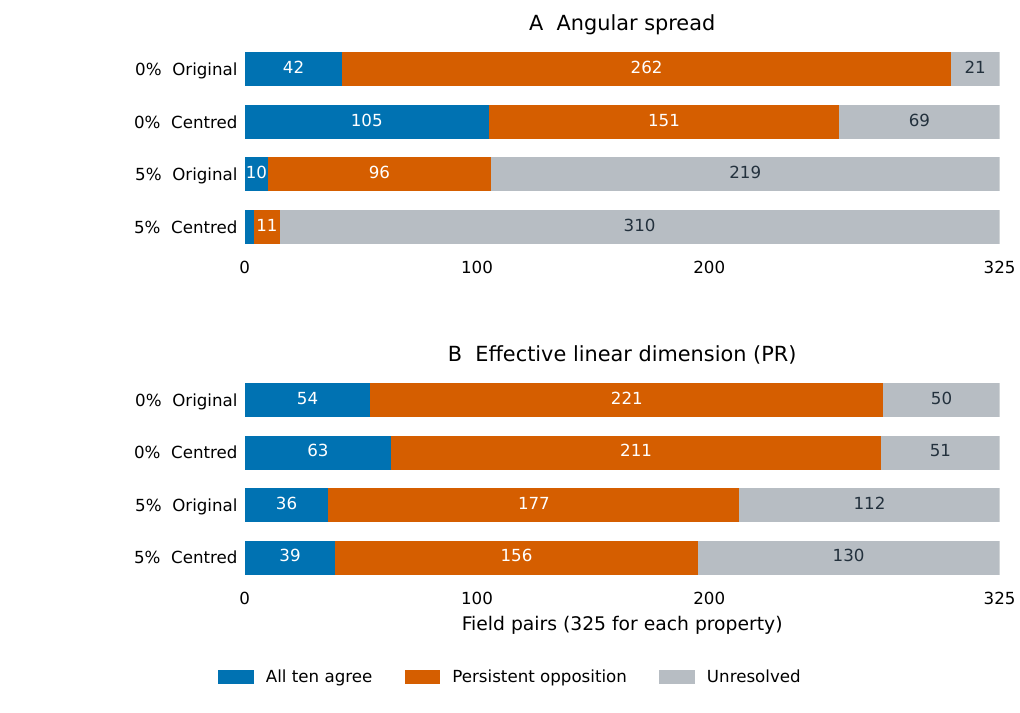}{How model agreement changes with the representation. Each bar covers all 325 pairs of areas. Blue: all ten models persistently give the same area the larger measured value. Orange: at least two fixed models persistently give opposite orderings. Grey: neither rule is met. Every ordering must hold across 26 conditions: the main 2,000 papers per area, twenty half samples of 1,000, and five further selections of 2,000. Original and centred representations use the same papers. Centring moves the origin to each model's mean over the balanced 52,000 unit vectors, then rescales the resulting vectors to length one. A cutoff of 5\% requires a difference greater than 5\% of the two area values' average in every condition; the 0 cutoff requires a nonzero difference. Table S20 reports all four cutoffs and whether the same models retain the same opposing area orderings.}{pairs}{1}

The alternative formulas often retain the disagreement, but not in every case. I count an opposition as retained only when at least one of the original opposing model pairs keeps the same area orderings. Alternative angle summaries retain 85.9\% of the original angular oppositions by this rule (225/262). Across the same 26 article conditions, entropy-based dimension retains 94.6\% of the original PR oppositions and D80 retains 86.0\%; 86.0\% survive both alternatives (190/221). These denominators are the original opposed pairs, not all area pairs. Because D80 returns whole numbers, it can give ties that contribute no persistent ordering.

The four word-level models do not account for all opposition either. Keeping only the six models trained for similarity still leaves opposite answers in 71.1\% of all 325 pairs for angular spread and 49.2\% for PR (Tables S14--S15).

Angular comparisons change much more when the common reference point moves to the mean position of the balanced 52,000 papers (Figure~\ref{fig:pairs}). After subtracting that mean and restoring unit vector length, opposition falls from 80.6\% to 46.5\% of all 325 pairs: from 262 to 151. But the new total does not tell us which original disagreements remain. Only 48.9\% of the original oppositions retain at least one of the same opposing model pairs with the same area orderings (128/262). Table S20 separates these from new oppositions or those supported by different model pairs. With the 5\% minimum-difference rule, only 11 opposed pairs remain instead of 96.

PR is less sensitive to this transformation. Opposed pairs fall from 221 to 211, and 92.3\% of the original oppositions retain at least one of the same opposing model pairs and area orderings (204/221). An angular comparison that repeats with other papers can therefore still depend strongly on the origin from which angles are measured.

The attempt to identify separate groups from links between nearby papers reached a different limit. Weak connections might seem to suggest an area divided into research topics. Yet artificial elongated clouds and isolated extreme points produce that signal without any thematic groups. It also changes with sample size and the rule for linking neighbours, with selection alerts in 51 of 260 cases. Supplement S6 retains this unsuccessful check because it shows a problem that repeatability alone cannot solve: a geometric signal may not describe the feature of research we want to study.

\FloatBarrier
\section{Discussion}
\subsection{What should be checked before using a map}
A high overall agreement score leaves some practical questions open. The models can share the broad arrangement of areas while directing a reader towards different papers or giving opposite comparisons between areas. Suppose one model gives area A a larger angular spread than B, and the result repeats with other articles. That shows stability to article selection within this model. If another model persistently reverses the order, more selections with the first model do not resolve the disagreement. The claim needs to be checked by changing the model as well as the articles.

The concern follows earlier work on method-dependent science maps \citep{refa0d2d707e6,ref6a2f5912bc} and complements assessments against external classifications and relationships \citep{refbe130e2973,refc0e2cc29e0}. Combining structural and neighbour comparisons also has precedent \citep{ref4e1e0d5582}.

Interpretation also depends on what the measure describes. Larger angles mean that vectors point further apart; a larger effective linear dimension means that their variation is less concentrated in a few directions. Those are descriptions of the representation. Calling an area more thematically diverse or interdisciplinary would require a further link between those numerical properties and the research itself. Related work on embedding geometry cautions against assigning a broad meaning to a convenient measure \citep{rudman-etal-2022-isoscore}. The failed fragmentation check illustrates why: even a geometric difference that repeats can support the wrong explanation.

For someone using these maps to compare research areas, the practical consequence is to check the comparison itself. An average agreement score describes what the representations share overall. Repeating article selection asks whether the answer depends on the papers chosen. Following the same opposing models adds a further check: whether a stable answer within each model transfers across representations. Reporting the size of the difference and the processing choices makes clear which of these claims the evidence supports. External evidence is still needed to interpret a geometric contrast as a property of the research itself.

\subsection{What the additional checks change}
The additional checks narrow the explanations for the main findings. The shared arrangement of area centres is not tied to one set of papers, to choosing 128 rather than 256 or 512, or to including one particular area. The low neighbour overlap is also not largely removed by the recorded quality filter: after excluding flagged and duplicate records, two models still share about eight papers in a list of 25. These checks strengthen those observations within the tested collection, without establishing the accuracy of the shared relationships.

For input text, the repeated selections preserve an average result that hides different model behaviours. Giving titles alone changes fewer neighbours than switching models for some models, and more for others. Replacing mean pooling with the first or last contextual output in the four word-level models also reverses the overall ordering. The practical consequence is to specify both what text a model reads and how its outputs become a paper vector. A statement that text and model choice matter equally would go beyond these results.

For angular spread, the common reference point is another consequential choice. Moving it to the mean of the balanced article set and rescaling the vectors changes many answers about which area has the larger spread. Repeating article selection does not remove this dependence or establish one representation as correct. PR changes less, although alternative dimension formulas still fail to retain every comparison. A claim about spread therefore needs to identify the representation, the measure and the size of the difference.

The discipline checks examine one concrete difference: models agree less about the relationships among Medicine papers than among Energy papers under the main rule (Supplement S5). This does not mean that Medicine is less coherent, and it does not always have the lowest neighbour overlap. The structural difference remains when the contributions of specialties within each area are balanced, and when the two biomedical models, BioBERT and PubMedBERT, are excluded. Neither tested explanation accounts for the whole difference. Models with related training also sometimes give similar results, but these apparent families change with the rule used to form paper vectors. With ten models that differ simultaneously in training texts, internal design and the task used to train them, these associations cannot identify one of those features as the cause.

The time comparison applies the same current models to papers from older and newer publication periods. Structural agreement increases slightly on average. Neighbour agreement falls by 0.87 percentage points when each group offers all its available papers, but rises by 0.91 points when every group offers 2,048 candidates. The apparent trend therefore depends on whether older and newer groups are allowed different search sizes. It describes agreement among present-day representations of those papers; it does not, by itself, show that science has historically converged.

\subsection{What remains open}
The collection requires English-labelled records with usable abstracts. Its reinforcement of smaller area--period groups helps comparison but changes proportions. Source errors remain, and cleaning is not a certification of every record. Repeated selections cannot repair missing coverage or make these results representative of all scientific work.

The ten models also set a boundary. Four were not trained specifically for document similarity. The six-model check shows that disagreement is not confined to those four, but does not make this set representative of all embedding models. To decide which area ordering or article relationship is better supported, we need evidence outside these vectors, such as independent assessments of the papers.

Several extensions and controls also followed earlier results. Recording their rules before the new calculations makes the decisions traceable, but does not turn them into predictions made before the study began. A useful next step would test a specific map-based claim against independent judgments in a research area. Adding models alone would not settle what their shared or conflicting geometry means.

\section{Conclusion}
Changing the embedding model can preserve the broad arrangement of science while changing which papers appear nearby and which area has more widely spread representations. Giving a model titles alone can also produce large changes, with effects that depend on the model and the rule used to form paper vectors. Choosing other articles tests one source of dependence; changing models or the reference point tests others. Before using a map to support a particular relationship, we should check that relationship under the choices that can affect it and show how large the changes are. The tested models' agreement tells us what their representations share. Deciding whether that relationship describes the research well still needs independent evidence.

\section*{Declarations}
\textbf{Author contributions.}
Alejandro Treny Ortega: Conceptualization; Methodology; Software; Validation; Formal analysis; Investigation; Data curation; Visualization; Writing -- original draft; Writing -- review \& editing.
\par\medskip
\textbf{Competing interests.}
The author declares no competing interests.
\par\medskip
\textbf{Funding.}
This research received no external funding.
\par\medskip
\textbf{Data and code availability.}
Code and analysis data (version 1.1.0) are publicly available on Zenodo at \url{https://doi.org/10.5281/zenodo.22876602} \citep{treny2026reproduction}. A \href{https://github.com/aleetreny/Shape-of-Science-Reproducibility}{GitHub mirror} provides the same files and commands. They preserve article identifiers, exact selections, measurements by model, counts by query and controls. The public route repeats the documented statistical comparisons from those frozen measurements; it excludes stored vectors, neighbour lists and model weights. Historical title/abstract tables are not redistributed because rights to the plaintext abstracts have not been established. Fresh OpenAlex retrieval may return different text, so an exact rerun from the historical inputs is not guaranteed. Code is MIT; original results and documentation are CC BY 4.0; OpenAlex metadata retain CC0. Supplement S8 details the tested scope.

\begingroup
\interlinepenalty=10000
\bibliographystyle{plainnat}
\bibliography{references,context_references}
\endgroup
\end{document}


\begin{center}
{\LARGE\bfseries Supplementary material\par}
\medskip
{\large How much does the map of science depend on the embedding model?\par}
\medskip
Alejandro Treny Ortega\par
University College London, London, United Kingdom\par
\smallskip
21 September 2026
\end{center}

This supplement provides definitions, methodological checks, and detailed result summaries. The accompanying reproducibility package (version 1.1.0), including the publicly released numerical data and code, is archived separately on Zenodo at \url{https://doi.org/10.5281/zenodo.22876602}. The package README and \texttt{SELECTION.json} document the included files.

\textbf{Which articles belong to which check?}
The 500,000 records are the starting collection. The text experiment uses 52,000 of them to compare title, abstract and both; its earlier 26,000-paper trial is included, not added. The geometric analysis in S6--S7 reuses that same set of 52,000. A different set of 52,000 is used for the shared-passage check in S1. The matched searches in S4 follow 10,850 starting papers while changing the papers available around them. These are overlapping uses of one collection, not separate downloads.

Here, Field means one of the 26 broad research areas and Subfield means a narrower specialty. The main article introduces the comparisons in that order; this supplement keeps the database names and technical definitions so that the calculations can be checked.

\medskip
{\small\setstretch{1}\setcounter{tocdepth}{1}%
\makeatletter
\patchcmd{\l@section}{1.0em \@plus\p@}{0.3em \@plus\p@}{}{\PackageError{supplement}{Unable to compact table of contents}{Review the section TOC spacing}}%
\makeatother
\tableofcontents}
\clearpage

\section{Corpus, models and provenance}
\subsection{Eligibility, quotas and quality flags}
OpenAlex's core catalogue was queried for 2000--2024 records of type article, review or conference-paper, with an English language label, an abstract, no retraction or paratext flag and a known primary Field. Local preparation required a non-empty title and at least 50 regex words in the reconstructed abstract. It preserved literal content rather than translating or generating missing summaries. The unit was a distinct OpenAlex work ID, not a deduplicated intellectual work.

Seeded random request blocks supplied the general cohort. Saved candidates were processed in their recorded order until 400,000 eligible, distinct IDs were retained. The 100,000-record supplement was assigned to the smallest of the 130 Field-period groups, with deterministic Field/year tie-breaking. This raised minimum coverage without reducing the larger groups. Exact year counts were retained rather than forcing equal quotas within each five-year period. Raw retrieval ended on 15 September 2026; API responses were collected over time, so the corpus is not a simultaneous database snapshot. Saved responses, ID order and content hashes are required to reconstruct it; seeds alone would not reproduce a changing API.

Cleaning rejected recognised non-abstract content, explicit withdrawal/retraction titles and clear language failures. For language, fastText's lid.176 model was applied to the whole abstract and three consecutive parts. A record was excluded only if the whole text had a non-English top label with score at least 0.9 and all three parts agreed on that language with scores at least 0.8. These are conservative operational scores, not calibrated probabilities or independent tests. Remaining uncertainty, possible mixed language, 50--79-word or unusually long abstracts, possible notices and duplicate DOI/text were marked. Content rules were developed by inspecting the downloaded corpus before embedding; they were not preregistered before extraction.

The final corpus retains 9,883 abstracts with language uncertainty, including 3,002 flagged as potentially mixed. These sets overlap. It also includes 42,849 abstracts of 50--79 words and 249 over 2,000 words. Subsequent analysis uses the flags for sensitivity rather than silently changing the corpus. Source checks establish agreement with downloaded records, not that every record is correct. Table S1 supplies the distribution and the full flag tables.
\clearpage
\par\noindent\begin{minipage}{\linewidth}
\fontsize{9.5}{11}\selectfont
\setlength{\tabcolsep}{4pt}
\renewcommand{\arraystretch}{1.03}
\captionof{table}{Clean corpus by Field, period and cohort.}\label{tab:S01}
\begin{tabular}{@{}>{\raggedright\arraybackslash}p{5.1cm}>{\raggedright\arraybackslash}p{1.23cm}>{\raggedright\arraybackslash}p{1.23cm}>{\raggedright\arraybackslash}p{1.23cm}>{\raggedright\arraybackslash}p{1.23cm}>{\raggedright\arraybackslash}p{1.23cm}>{\raggedright\arraybackslash}p{1.23cm}>{\raggedright\arraybackslash}p{1.23cm}@{}}
\toprule
\textbf{Field} & \textbf{2000-04} & \textbf{2005-09} & \textbf{2010-14} & \textbf{2015-19} & \textbf{2020-24} & \textbf{Base} & \textbf{Extra} \\
\midrule
Agricultural and Biological Sciences & 2,166 & 2,528 & 3,940 & 4,298 & 6,491 & 18,920 & 503 \\
Arts and Humanities & 2,166 & 2,166 & 2,186 & 2,732 & 3,380 & 10,696 & 1,934 \\
Biochemistry, Genetics and Molecular Biology & 2,739 & 4,071 & 5,251 & 5,596 & 6,840 & 24,497 & 0 \\
Business, Management and Accounting & 2,166 & 2,166 & 2,186 & 2,660 & 4,040 & 11,137 & 2,081 \\
Chemical Engineering & 2,166 & 2,166 & 2,166 & 2,166 & 2,166 & 1,195 & 9,635 \\
Chemistry & 2,166 & 2,166 & 2,166 & 2,166 & 2,166 & 7,290 & 3,540 \\
Computer Science & 3,310 & 5,594 & 7,094 & 6,858 & 9,972 & 32,828 & 0 \\
Decision Sciences & 2,166 & 2,166 & 2,166 & 2,166 & 2,166 & 3,817 & 7,013 \\
Earth and Planetary Sciences & 2,166 & 2,166 & 2,166 & 2,166 & 2,166 & 7,256 & 3,574 \\
Economics, Econometrics and Finance & 2,166 & 2,166 & 2,166 & 2,166 & 2,845 & 8,620 & 2,889 \\
Energy & 2,166 & 2,166 & 2,166 & 2,166 & 2,166 & 2,617 & 8,213 \\
Engineering & 6,990 & 11,310 & 15,203 & 14,738 & 18,907 & 67,148 & 0 \\
Environmental Science & 2,166 & 2,876 & 4,005 & 4,018 & 5,919 & 18,569 & 415 \\
Immunology and Microbiology & 2,166 & 2,166 & 2,166 & 2,166 & 2,166 & 4,456 & 6,374 \\
Materials Science & 2,166 & 2,166 & 2,962 & 3,152 & 3,787 & 13,335 & 898 \\
Mathematics & 2,166 & 2,166 & 2,166 & 2,166 & 2,166 & 4,581 & 6,249 \\
Medicine & 7,128 & 10,946 & 15,771 & 18,629 & 26,195 & 78,669 & 0 \\
Neuroscience & 2,166 & 2,166 & 2,166 & 2,166 & 2,166 & 6,824 & 4,006 \\
Nursing & 2,166 & 2,166 & 2,166 & 2,166 & 2,166 & 2,077 & 8,753 \\
Pharmacology, Toxicology and Pharmaceutics & 2,166 & 2,166 & 2,166 & 2,166 & 2,166 & 1,715 & 9,115 \\
Physics and Astronomy & 2,166 & 2,704 & 3,094 & 3,388 & 3,470 & 14,772 & 50 \\
Psychology & 2,166 & 2,166 & 2,166 & 2,577 & 3,970 & 10,816 & 2,229 \\
Social Sciences & 2,776 & 4,893 & 7,110 & 8,891 & 13,879 & 37,549 & 0 \\
Veterinary & 2,166 & 2,166 & 2,165 & 2,165 & 2,165 & 642 & 10,185 \\
Dentistry & 2,165 & 2,165 & 2,165 & 2,165 & 2,165 & 2,238 & 8,587 \\
Health Professions & 2,165 & 2,165 & 2,165 & 2,165 & 2,833 & 7,736 & 3,757 \\
\textbf{Total} & 68,427 & 83,908 & 101,289 & 107,858 & 138,518 & 400,000 & 100,000 \\
\bottomrule
\end{tabular}
\end{minipage}\par
\par\smallskip{\small Total: 500,000 distinct OpenAlex work IDs. Period columns contain base plus extra; the last two columns divide those same articles and must not be added again. Core catalogue; years 2000--2024; article, review and conference-paper; OpenAlex English label, non-empty title, reconstructible abstract with at least 50 regex words, known primary Field, no retraction or paratext flag. Additional cleaning rejects clear language/content failures and marks ambiguous cases. No claim that every retained abstract is exclusively English or correctly labelled. Exact annual/cohort counts and quality flags are included as CSVs; raw retrieval ended on 15 September 2026. The database was not a simultaneous snapshot.\par}

\FloatBarrier
\subsection{Checkpoints, input and truncation}
All ten models use the same ordered IDs and original title/abstract text. The customary separators, full model revisions and any adapter revisions are listed in Table S2. Outputs were saved as 32-bit vectors; subsequent geometric calculations use 64-bit arithmetic. Missing or degenerate outputs were checked, and each run records code, configuration, file and environment hashes.

Mean pooling averages every non-padding last-layer position, including model-added special tokens. CLS is the first-position state before any pooler; SEP is the final non-padding position, rather than the separator between title and abstract. Only BERT, SciBERT, BioBERT and PubMedBERT change under the CLS/SEP scenarios. SimCSE uses its pre-pooler CLS state; SPECTER2 uses its proximity adapter. MPNet and MiniLM retain their sentence-transformer mean-pooling configuration. Readout scenarios are not additional independently trained models.

The common-fragment comparison finds a shared prefix of content characters that fits every model, keeping each tokenizer and its customary separators. It uses a separate 52,000-ID panel and compares that fragment with the native representation of those same IDs. It does not equate token sequences between vocabularies. MiniLM's separate 512-token check uses its supported architectural limit; vectors for texts unaffected by the original truncation are reused exactly. Neither check demonstrates that longer input gives better thematic representations.
\clearpage
\par\noindent\begin{minipage}{\linewidth}
\small
\setlength{\tabcolsep}{4pt}
\renewcommand{\arraystretch}{1.03}
\captionof{table}{Frozen checkpoints, text formats and native truncation.}\label{tab:S02}
\begin{tabular}{@{}>{\raggedright\arraybackslash}p{2.8cm}>{\raggedright\arraybackslash}p{8.1cm}>{\raggedright\arraybackslash}p{4.25cm}@{}}
\toprule
\textbf{Model} & \textbf{Checkpoint and full revision} & \textbf{Input / truncation} \\
\midrule
SPECTER & {\footnotesize \nolinkurl{allenai/specter}\newline Revision: {\ttfamily\seqsplit{81cbfb43d4fc2e728d5b4201ce14987db8d0854c}}} & Title [SEP] abstract\newline Limit: 512\newline Truncated: 5.66\% \\
SPECTER2 & {\footnotesize \nolinkurl{allenai/specter2_base}\newline Revision: {\ttfamily\seqsplit{3447645e1def9117997203454fa4495937bfbd83}}\newline Adapter: \nolinkurl{allenai/specter2}\newline{\ttfamily\seqsplit{2081559630a80fc5851d8f798a05ba81e9468089}}} & Title [SEP] abstract\newline Limit: 512\newline Truncated: 5.00\% \\
SciNCL & {\footnotesize \nolinkurl{malteos/scincl}\newline Revision: {\ttfamily\seqsplit{ebc5348d184ba2fc9beee69b4e394263fce57b2e}}} & Title [SEP] abstract\newline Limit: 512\newline Truncated: 5.00\% \\
SciBERT & {\footnotesize \nolinkurl{allenai/scibert_scivocab_uncased}\newline Revision: {\ttfamily\seqsplit{24f92d32b1bfb0bcaf9ab193ff3ad01e87732fc1}}} & Title, blank line, abstract\newline Limit: 512\newline Truncated: 4.96\% \\
BERT & {\footnotesize \nolinkurl{google-bert/bert-base-uncased}\newline Revision: {\ttfamily\seqsplit{86b5e0934494bd15c9632b12f734a8a67f723594}}} & Title, blank line, abstract\newline Limit: 512\newline Truncated: 6.87\% \\
MPNet & {\footnotesize \nolinkurl{sentence-transformers/all-mpnet-base-v2}\newline Revision: {\ttfamily\seqsplit{e8c3b32edf5434bc2275fc9bab85f82640a19130}}} & Title, blank line, abstract\newline Limit: 384\newline Truncated: 20.16\% \\
MiniLM & {\footnotesize \nolinkurl{sentence-transformers/all-MiniLM-L6-v2}\newline Revision: {\ttfamily\seqsplit{1110a243fdf4706b3f48f1d95db1a4f5529b4d41}}} & Title, blank line, abstract\newline Limit: 256\newline Truncated: 49.58\% \\
PubMedBERT & {\footnotesize \nolinkurl{microsoft/BiomedNLP-BiomedBERT-base-uncased-abstract-fulltext}\newline Revision: {\ttfamily\seqsplit{e1354b7a3a09615f6aba48dfad4b7a613eef7062}}} & Title, blank line, abstract\newline Limit: 512\newline Truncated: 4.75\% \\
BioBERT & {\footnotesize \nolinkurl{dmis-lab/biobert-v1.1}\newline Revision: {\ttfamily\seqsplit{551ca18efd7f052c8dfa0b01c94c2a8e68bc5488}}} & Title, blank line, abstract\newline Limit: 512\newline Truncated: 9.00\% \\
SimCSE & {\footnotesize \nolinkurl{princeton-nlp/unsup-simcse-bert-base-uncased}\newline Revision: {\ttfamily\seqsplit{218baf6bc48c463a53f5b4e1c568299537fab124}}} & Title, blank line, abstract\newline Limit: 512\newline Truncated: 6.87\% \\
\bottomrule
\end{tabular}
\end{minipage}\par
\par\smallskip{\small Truncation is measured on the full 500,000-article title-plus-abstract corpus and is not an error rate. The principal MiniLM limit remains 256 tokens; its 512-token variant is a separate sensitivity. CLS/SEP alternatives apply only to BERT, SciBERT, BioBERT and PubMedBERT. SEP takes the last non-padding position, not the title separator. Tokenizer differences mean that equal token limits would not imply equal text. Model-card URLs and all revisions are in the CSV.\par}

\FloatBarrier

\section{Agreement measures and sampling stability}
This section explains how agreement between two models is calculated, then checks how much the result changes when other articles are selected. Those are two different questions.

\subsection{Coverage and chronology}
\par\noindent\begin{minipage}{\linewidth}
\small
\setlength{\tabcolsep}{4pt}
\renewcommand{\arraystretch}{1.13}
\captionof{table}{Coverage and chronology of the comparison blocks.}\label{tab:S03}
\begin{tabular}{@{}>{\raggedright\arraybackslash}p{2.6cm}>{\raggedright\arraybackslash}p{4.1cm}>{\raggedright\arraybackslash}p{3.4cm}>{\raggedright\arraybackslash}p{5.05cm}@{}}
\toprule
\textbf{Block} & \textbf{Coverage} & \textbf{When specified} & \textbf{Interpretation} \\
\midrule
Initial comparison & 500,000 corpus; 130 cells; 45 fixed model pairs & Before this phase's agreements & Normalised vectors; corrected CKA, k25; CLS/SEP and size checks. \\
Common fragment & 52,000 paired IDs, distinct from the input panel & Before first scientific comparison & Same content characters; native tokenizers and separators. \\
Centres / random groups & 256 per centre; 26/217/26 centres & Added after initial results & Matched dates/sizes; addresses centre averaging, not thematic validity. \\
Input expansion & 26,000 to 52,000 nested IDs & After pilot; before expanded result & Three inputs and pooling crossed; 100 comparable selections. \\
Matched local searches & 217 groups, 50 queries, ten selections & Expanded phase design & 256 candidates per scope, including identical queries and matched dates. \\
Atlas & 217 Subfields; 183 in sensitivity panel; 12 paper examples & After global results; before reading selected titles & Rule-selected examples; source errors retained. \\
Morphology & 52,000 IDs; alternative measures and synthetic clouds & Exploratory extension after core results & Protocol before new real-data metrics. Neighbour-fraction check added during execution. \\
Field pairs & 325 pairs per property; 26 saved conditions & After morphology results & Direction/magnitude rules fixed before new pair counts; no new inference. \\
\bottomrule
\end{tabular}
\end{minipage}\par
\par\smallskip{\small This is a sequence of documented phases, not a preregistered study. Sources: ANALYSIS\_PROTOCOL, ROBUSTNESS\_PROTOCOL, CASE\_ATLAS\_PROTOCOL, MORPHOLOGY\_PROTOCOL and FIELD\_PAIR\_PROTOCOL, with their dated decision entries. Later controls are not relabelled as initial hypotheses.\par}

\subsection{Alignment and neighbour definitions}
For each comparison, rows of $X$ and $Y$ identify the same articles. Vectors are scaled to unit length, and columns are then centred for CKA and Procrustes. Let $n$ be the row count and define
\begin{align*}
S_{XY}&=\lVert X^\top Y\rVert_F^2, &
R_{XY}&=\sum_i\lVert X_i\rVert^2\lVert Y_i\rVert^2,\\
T_X&=\sum_i\lVert X_i\rVert^2, &
H_{XY}&=S_{XY}-\frac{n}{n-2}R_{XY}
+\frac{T_XT_Y}{(n-1)(n-2)}.
\end{align*}
The reported corrected alignment is
\[
\mathrm{CKA}_{\mathrm{corr}}(X,Y)=\frac{H_{XY}}{\sqrt{H_{XX}H_{YY}}}.
\]
The common HSIC factor cancels. Negative corrected values are retained, and degenerate inputs are rejected. The implementation was checked independently using relation matrices with zero diagonals. The correction addresses finite-sample inflation \citep{pmlr-v97-kornblith19a,refa961d84a4e}; the complete ratio is not asserted to be exactly unbiased.

Procrustes similarity is $\lVert X^\top Y\rVert_*/\sqrt{T_XT_Y}$, where the numerator is the sum of singular values. Its arccosine gives the reported angular distance. Rotation/reflection and a global scale can be aligned; an arbitrary deformation cannot. Unequal embedding dimensions can be understood by adding zero coordinates to the smaller representation. Normalisation precedes centring, so this procedure does not imply invariance to translating the original vectors and normalising them again \citep{NEURIPS2021_252a3dba}.

The distance-rank check correlates cosine distances for 20,000 distinct article pairs selected with a recorded seed, identically in both representations. Ties receive average ranks. These pair observations and the alternative alignment measures are not treated as independent replications \citep{pmlr-v285-williams24a}.

For query $i$ and models $a,b$, neighbour overlap is
\[
O_k(i;a,b)=\frac{|N_k^a(i)\cap N_k^b(i)|}{k}.
\]
The query is excluded by ID. In the core exact search, similarities are computed in 64-bit blocks and rounded to twelve decimal places for deterministic tie resolution by global row order. Independent sorting checks verify the stored lists. The combinatorial reference for independent uniform lists among $M$ eligible candidates is $k/M$; the adjusted overlap is $(O_k-k/M)/(1-k/M)$. This reference does not remove density, popular-neighbour or composition effects. The primary results therefore retain raw overlap alongside explicitly matched candidate pools.

\subsection{Selection screens and retained alerts}
The original screen uses twenty selections at each of 1,024 and 2,048 articles per Field-period group, paired between models. It flags an absolute median shift above 0.02 or a central 95\% selection width above 0.04 at the larger size. A row permutation per model/group provides an alignment diagnostic rather than a permutation-test $p$-value. The quality sensitivity retains one representative per connected duplicate DOI/text component and excludes specified flags, leaving 448,886 records. It changes the operational population and does not replace the main corpus.

Subfield screens use twenty nested selections up to 256/512 articles, subject to availability. Groups below eight articles lack a two-size diagnosis. If the larger draw equals the full group, its zero selection width is explicitly marked and cannot establish precision for a wider population. Table S4 keeps these cases separate from the original Field-period alerts.
\par\noindent\begin{minipage}{\linewidth}
\small
\setlength{\tabcolsep}{4pt}
\renewcommand{\arraystretch}{1.13}
\captionof{table}{Agreement of alternative measures and retained stability alerts.}\label{tab:S04}
\begin{tabular}{@{}>{\raggedright\arraybackslash}p{6.1cm}>{\raggedright\arraybackslash}p{1.2cm}>{\raggedright\arraybackslash}p{3.0cm}>{\raggedright\arraybackslash}p{2.1cm}>{\raggedright\arraybackslash}p{2.1cm}@{}}
\toprule
\textbf{Measures compared} & \textbf{Cells} & \textbf{Mean rank agreement} & \textbf{Minimum} & \textbf{Maximum} \\
\midrule
Corrected CKA / Procrustes & 130 & 0.891 & 0.793 & 0.960 \\
Corrected CKA / Distance ranks & 130 & 0.881 & 0.777 & 0.963 \\
Procrustes / Distance ranks & 130 & 0.789 & 0.649 & 0.893 \\
\bottomrule
\end{tabular}
\end{minipage}\par
\par\smallskip{\small The top panel reports descriptive Spearman correlations of model-pair agreements within each of 130 Field-period cells. These correlations compare measures; they do not validate thematic content. All 41 saved summaries and all individual original alerts accompany the table.\par}
\begin{itemize}
\item Original Field-period screen: 50 alerts among 5,850 comparisons; largest central-95\% selection width among flagged cases: 0.064.
\item The larger sample does not equal the entire group: 2,628 alerts / 8,235 comparisons; 183 Subfields.
\item The larger sample equals the entire group: 213 alerts / 3,060 comparisons; 68 Subfields.
\end{itemize}
\par\smallskip{\small A zero selection width when the whole group is used does not establish stability at a smaller size. These are within-corpus screens, not population confidence intervals. Alerts from different designs remain separate.\par}

\FloatBarrier

\section{Text input, pooling and truncation}
This experiment keeps the papers fixed and changes what each model reads. The main question is whether that change matters more or less than choosing another model.

\subsection{Model-specific input effects}
The title/abstract experiment uses 52,000 articles: 400 per Field-period group and 2,000 per Field, selected by a fixed ID-based order. The first trial used 26,000; all those IDs remain in the larger set. For each model, the model-change effect is one minus mean agreement with the other nine models on title plus abstract. The input-change effect is one minus agreement between that model's combined input and its title-only or abstract-only input. Their difference is positive when model change is larger. Neighbour changes use the same 2,000 candidates per Field; self is excluded.

A symmetric summary of input transitions checks dependence on the combined-input reference. Mean, CLS and SEP scenarios cross all three inputs; only the four word-level BERTs change readout, which also changes the cross-model reference. Saved combined-input and pilot vectors were reused with exact ID/text alignment and bitwise checks. The analysis makes a practical comparison of input conditions, not a causal decomposition of architecture and content.
\par\noindent\begin{minipage}{\linewidth}
\small
\setlength{\tabcolsep}{4pt}
\renewcommand{\arraystretch}{1.13}
\captionof{table}{Input changes by model, averaged equally over the 26 Fields.}\label{tab:S05}
\begin{tabular}{@{}>{\raggedright\arraybackslash}p{3.0cm}>{\raggedright\arraybackslash}p{2.7cm}>{\raggedright\arraybackslash}p{2.7cm}>{\raggedright\arraybackslash}p{3.1cm}>{\raggedright\arraybackslash}p{3.1cm}@{}}
\toprule
\textbf{Model} & \textbf{Title: shape} & \textbf{Abstract: shape} & \textbf{Title: neighbours} & \textbf{Abstract: neighbours} \\
\midrule
SPECTER & 0.208 & 0.044 & 61.1\% & 31.3\% \\
SPECTER2 & 0.203 & 0.048 & 60.6\% & 31.3\% \\
SciNCL & 0.205 & 0.028 & 57.3\% & 23.9\% \\
SciBERT & 0.532 & 0.014 & 84.2\% & 15.8\% \\
BERT & 0.418 & 0.013 & 83.6\% & 18.0\% \\
MPNet & 0.189 & 0.022 & 51.7\% & 17.7\% \\
MiniLM & 0.246 & 0.031 & 58.1\% & 21.5\% \\
PubMedBERT & 0.526 & 0.017 & 82.9\% & 16.2\% \\
BioBERT & 0.636 & 0.024 & 83.4\% & 19.2\% \\
SimCSE & 0.496 & 0.028 & 83.8\% & 26.3\% \\
\bottomrule
\end{tabular}
\end{minipage}\par
\par\smallskip{\small Same 52,000 IDs and 2,000 candidates per Field. Shape change: 1 minus corrected CKA. Neighbour change: fraction of the 25 neighbours replaced. Title/abstract conditions compare each model with its own full input. Model-choice means and all Field-specific values are included as CSVs, alongside the full recipe cross. These are practical input changes with native text limits, not a causal separation of encoder and text length.\par}
\par\smallskip{\small Title-only model-minus-input contrast (positive: changing model has the larger mean effect): MEAN, shape = 0.0042; MEAN, neighbours = -0.0191; CLS, shape = 0.0363; CLS, neighbours = 0.0304; SEP, shape = 0.0627; SEP, neighbours = 0.0171. Only the four word BERTs change readout in CLS/SEP. Close means do not establish equivalence.\par}

\FloatBarrier
\subsection{The 26k to 52k stability check}
Both sizes were assessed with 20 and then 100 selections without replacement, using 500 articles per Field in the pilot and 1,000 in the expansion, balanced across dates. The first twenty repeats reproduce the earlier diagnostic. The same 0.02 median-shift and 0.04 central-width rules apply to the principal shape contrast. Retaining the original 31 keys distinguishes persistent, resolved and newly appearing alerts. The 100-selection comparison yields 72 to five alerts, while the historical twenty-selection comparison yields 31 to four. These counts must not be substituted for one another. They do not extend automatically to every readout or neighbour analysis.
\par\noindent\begin{minipage}{\linewidth}
\small
\setlength{\tabcolsep}{4pt}
\renewcommand{\arraystretch}{1.13}
\captionof{table}{Input-alert transitions for the nested 26k and 52k panels.}\label{tab:S06}
\begin{tabular}{@{}>{\raggedright\arraybackslash}p{3.5cm}>{\raggedright\arraybackslash}p{1.95cm}>{\raggedright\arraybackslash}p{1.7cm}>{\raggedright\arraybackslash}p{1.7cm}>{\raggedright\arraybackslash}p{1.45cm}>{\raggedright\arraybackslash}p{1.95cm}>{\raggedright\arraybackslash}p{1.2cm}@{}}
\toprule
\textbf{Selections / scope} & \textbf{Contrasts} & \textbf{26k alerts} & \textbf{52k alerts} & \textbf{Persist} & \textbf{Resolved} & \textbf{New} \\
\midrule
20 / model-Field & 520 & 31 & 4 & 3 & 28 & 1 \\
20 / Field average & 52 & 0 & 0 & 0 & 0 & 0 \\
100 / model-Field & 520 & 72 & 5 & 5 & 67 & 0 \\
100 / Field average & 52 & 0 & 0 & 0 & 0 & 0 \\
\bottomrule
\end{tabular}
\end{minipage}\par
\par\smallskip{\small The five remaining model-Field-input alerts with 100 selections are: BioBERT / Agricultural and Biological Sciences / abstract (width 0.047); BioBERT / Earth and Planetary Sciences / abstract (width 0.045); BioBERT / Economics, Econometrics and Finance / abstract (width 0.042); BioBERT / Immunology and Microbiology / abstract (width 0.063); PubMedBERT / Health Professions / title (width 0.040). No alert does not imply an effect different from zero. Original 20-selection and later 100-selection counts are kept separate.\par}

\PanelFigure{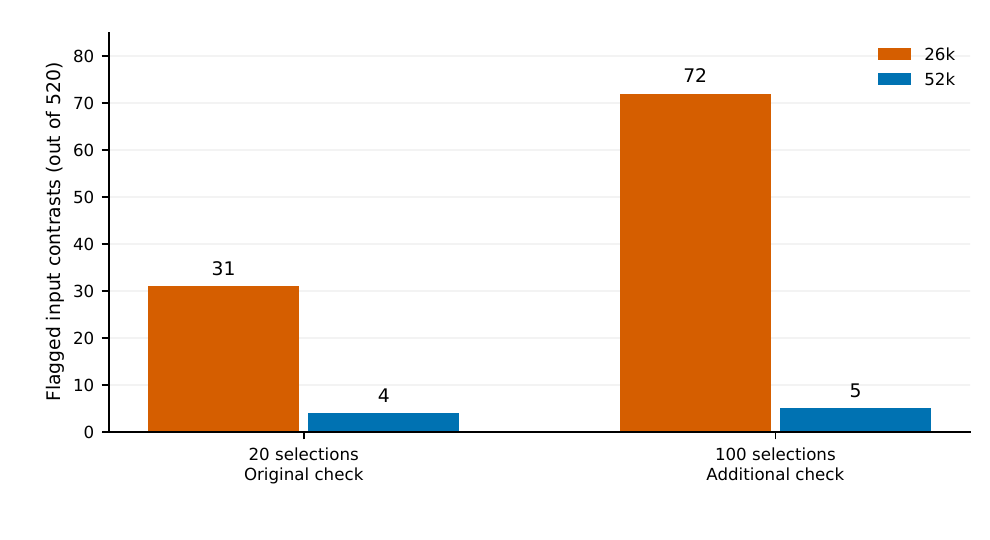}{Input stability alerts for nested samples. Counts are model--Field--input contrasts out of 520. The original 20-selection comparison and later 100-selection check use the same recorded thresholds, but their counts are not interchangeable. These are selection-sensitivity checks within the fixed corpus. The 52 Field-averaged contrasts are reported separately in Table S6.}{input-alerts}{1}
\FloatBarrier

\section{Scale, neighbourhoods and concrete cases}
These checks move from broad areas to specialties and individual papers. They separate a change in search scope from a change in the number of available papers.

\subsection{Matched scopes and candidate sizes}
Within-group comparisons use 128, 256 and 512 articles, with nested ID-based selections and the same 183 eligible Subfields at all three sizes. The native analysis retains all 252 Subfields where the measure is defined; groups with exactly $k+1$ articles have uninformative complete neighbour lists and are marked accordingly. A separate time comparison retains the same 125 Subfields with at least 128 articles in every period.

Centre comparisons average unit article vectors and scale each centre to unit length before corrected CKA. Direct and weighted-Subfield reconstructions of Field centres were checked for equality. The original matched centres contain 256 articles per group. Its twenty restricted selections change which Subfield represents each Field, not the articles within those Subfields. The repeated centre, size, matched-random and omission analyses now reported in the main text are specified in S9. Random references retain cross-model article correspondence; they are not null tests of independence.

The paired search fixes 50 queries per eligible Subfield, with 256 candidates in each scope and ten selections. The same 50 queries are included in both candidate sets; only the other 206 positions change scope. Date counts match. One query per model, selection, group and scope was independently sorted, giving 43,400 checks. Results for those 10,850 queries have repeated candidate controls; local scores for all 500,000 records do not.
\par\noindent\begin{minipage}{\linewidth}
\small
\setlength{\tabcolsep}{4pt}
\renewcommand{\arraystretch}{1.13}
\captionof{table}{Matched Field/Subfield neighbour searches.}\label{tab:S07}
\begin{tabular}{@{}>{\raggedright\arraybackslash}p{0.65cm}>{\raggedright\arraybackslash}p{2.7cm}>{\raggedright\arraybackslash}p{2.7cm}>{\raggedright\arraybackslash}p{3.0cm}>{\raggedright\arraybackslash}p{3.3cm}>{\raggedright\arraybackslash}p{2.1cm}@{}}
\toprule
\textbf{k} & \textbf{Field overlap} & \textbf{Subfield overlap} & \textbf{Difference (pp)} & \textbf{Lower / higher} & \textbf{Subfields} \\
\midrule
10 & 38.7\% & 36.9\% & -1.73 & 145 / 72 & 217 \\
25 & 46.7\% & 45.3\% & -1.35 & 127 / 90 & 217 \\
50 & 54.5\% & 53.7\% & -0.79 & 109 / 108 & 217 \\
\bottomrule
\end{tabular}
\end{minipage}\par
\par\smallskip{\small Difference is Subfield minus Field. Each scope contains 256 candidates, including the same 50 query articles, with matched dates; self is excluded during search. Means use ten candidate selections. The broader Field comparison is conditioned on the queries and is not 256 unrestricted Field draws. All 217 Subfields appear in the paired design; 183 meet all size conditions of the separate 128/256/512 sensitivity panel. Full size/readout/k results accompany the table.\par}

\FloatBarrier
\subsection{Subfield persistence}
The atlas compares ranks among the same 183 Subfields under 27 combinations of three candidate sizes, three readouts and three neighbour counts. Example rules were fixed after the overall results but before reading selected article titles. The six upper and six lower examples are descriptive extremes under those rules. Their rank ranges locate persistent relative agreement, not semantically correct or incorrect Subfields.
\PanelFigure{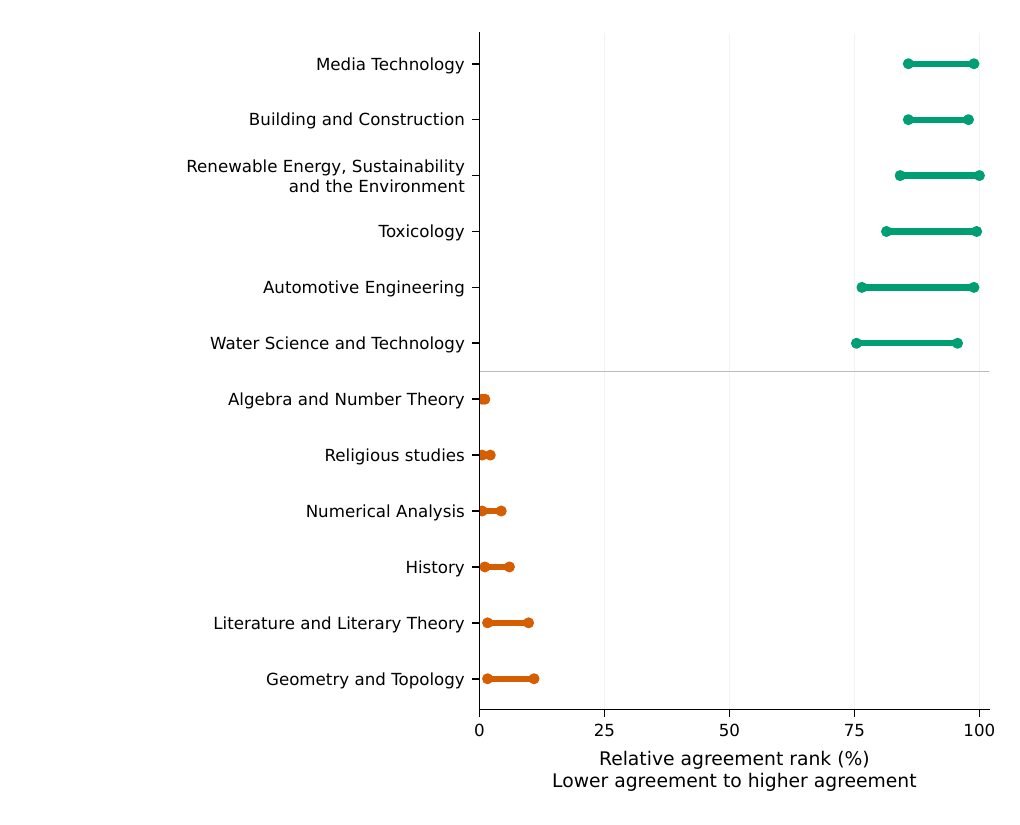}{Relative neighbourhood-agreement ranks for twelve rule-selected Subfield examples. Ranges span 27 combinations of candidate size, readout and neighbour count, using the same 183 eligible Subfields. The upper/lower six illustrate the recorded higher/lower agreement rule. These are sensitivity ranges, not confidence intervals; higher agreement does not imply better thematic accuracy.}{subfields}{1}
\FloatBarrier
\subsection{Article and relation examples}
The article atlas uses ten candidate selections for the fixed queries. Link comparisons are restricted to 531,650 directed relations whose endpoints remain eligible, avoiding the error of treating an unavailable target as a rejected neighbour. For each displayed source, both persistent and model-dependent targets are retained in the accompanying files. Figure S3 selects the eligible relation with maximal model disagreement for that source; it cannot estimate how common disagreement is.

The examples were inspected against saved OpenAlex records. The history-book, infection/politics and announcement cases remain visible, as does one target with an incomplete abstract. A diagnostic for 82 generic titles removes five fixed queries and barely changes the overall query mean (45.344\% to 45.349\%). That narrow check neither estimates the prevalence of thematic classification errors nor cleans all candidates.
{\small
\setlength{\tabcolsep}{4pt}
\renewcommand{\arraystretch}{1.16}
\begin{longtable}{@{}>{\raggedright\arraybackslash}p{4.9cm}>{\raggedright\arraybackslash}p{3.1cm}>{\raggedright\arraybackslash}p{1.25cm}>{\raggedright\arraybackslash}p{2.0cm}>{\raggedright\arraybackslash}p{3.55cm}@{}}
\caption{All twelve rule-selected article examples, including source caveats.}\label{tab:S08}\\
\toprule
\textbf{Article and OpenAlex ID} & \textbf{Subfield label} & \textbf{Mean} & \textbf{Selection range} & \textbf{Inspection note} \\
\midrule
\endfirsthead
\multicolumn{5}{l}{\small\itshape Table \thetable{} (continued)} \\
\toprule
\textbf{Article and OpenAlex ID} & \textbf{Subfield label} & \textbf{Mean} & \textbf{Selection range} & \textbf{Inspection note} \\
\midrule
\endhead
\midrule
\multicolumn{5}{r}{\small Continued on next page} \\
\endfoot
\bottomrule
\endlastfoot
SIRT6 overexpression induces massive apoptosis in cancer cells but not in normal cells\newline{\footnotesize\nolinkurl{W2022976570}; 2011} & Geriatrics and Gerontology & 79.3\% & 76.7\%--82.4\% & No clear issue in this limited inspection; not expert validation. \\
Comparison of The Effectiveness of Scientific Approach and Problem-Solving Approach in Problem-Based Learning in Class IX of SMP Negeri 3 Pangsid\newline{\footnotesize\nolinkurl{W2998141762}; 2019} & Applied Mathematics & 84.1\% & 75.1\%--87.4\% & Mathematics education under Applied Mathematics; administrative label. \\
A systematic analysis on the managerial non-efficiency of the state-owned businesses\newline{\footnotesize\nolinkurl{W2359252036}; 2002} & General Decision Sciences & 76.9\% & 74.7\%--78.8\% & No clear issue in this limited inspection; not expert validation. \\
Application of multi-agent approach to development of the petrol dispatching system of gas stations network\newline{\footnotesize\nolinkurl{W3107610255}; 2020} & Discrete Mathematics and Combinatorics & 78.2\% & 74.0\%--82.4\% & No clear issue in this limited inspection; not expert validation. \\
Persuasive Talk, Delegation, and Announcements\newline{\footnotesize\nolinkurl{W2252942328}; 2012} & Safety Research & 78.2\% & 72.1\%--83.3\% & Persuasion economics under Safety Research; label caveat. \\
Road risk-perception and pedestrian injuries among students at Ain Shams University, Cairo, Egypt\newline{\footnotesize\nolinkurl{W2081778888}; 2011} & Safety, Risk, Reliability and Quality & 76.4\% & 70.8\%--85.2\% & No clear issue in this limited inspection; not expert validation. \\
Invasion Strategy\newline{\footnotesize\nolinkurl{W2011180849}; 2003} & Political Science and International Relations & 18.7\% & 16.1\%--21.0\% & Bacterial-infection abstract labelled as politics. \\
Beyond the witch trials: Witchcraft and magic in Enlightenment Europe\newline{\footnotesize\nolinkurl{W7045581866}; 2004} & Nuclear and High Energy Physics & 21.4\% & 19.2\%--23.7\% & History-book description labelled as physics; type caveat. \\
Optical properties of cobalt clusters implanted in thin silica layers\newline{\footnotesize\nolinkurl{W1966552042}; 2006} & Ceramics and Composites & 23.2\% & 21.3\%--24.4\% & Substantial formula code in abstract; input caveat. \\
Announcement\newline{\footnotesize\nolinkurl{W4234898602}; 2010} & Ophthalmology & 22.2\% & 19.9\%--25.8\% & Announcement / atlas review; not detected by the original notice screen. \\
Conversion of a rule based to an object oriented expert system\newline{\footnotesize\nolinkurl{W2117277818}; 2002} & Software & 25.0\% & 22.8\%--26.0\% & No clear issue in this limited inspection; not expert validation. \\
`Thare is na leid on life of lordship hym like': linguistic means of depicting king arthur in the poem Golagros and Gawain\newline{\footnotesize\nolinkurl{W3110667008}; 2020} & Classics & 25.5\% & 24.4\%--26.3\% & No clear issue in this limited inspection; not expert validation. \\
\end{longtable}
}
\par\smallskip{\small Numbers are k25 overlap averaged over 45 model pairs; ranges span ten candidate selections. The first/last six examples come from high/low agreement sources under the recorded rule. Titles were read after selection. Full source/target IDs, titles and per-model supports for both relation types are in the accompanying CSVs. The most model-dependent link was selected for each source; this does not estimate prevalence. An incomplete target abstract is also retained and marked in Figure S3. None of these checks validates OpenAlex labels.\par}

\PanelFigure{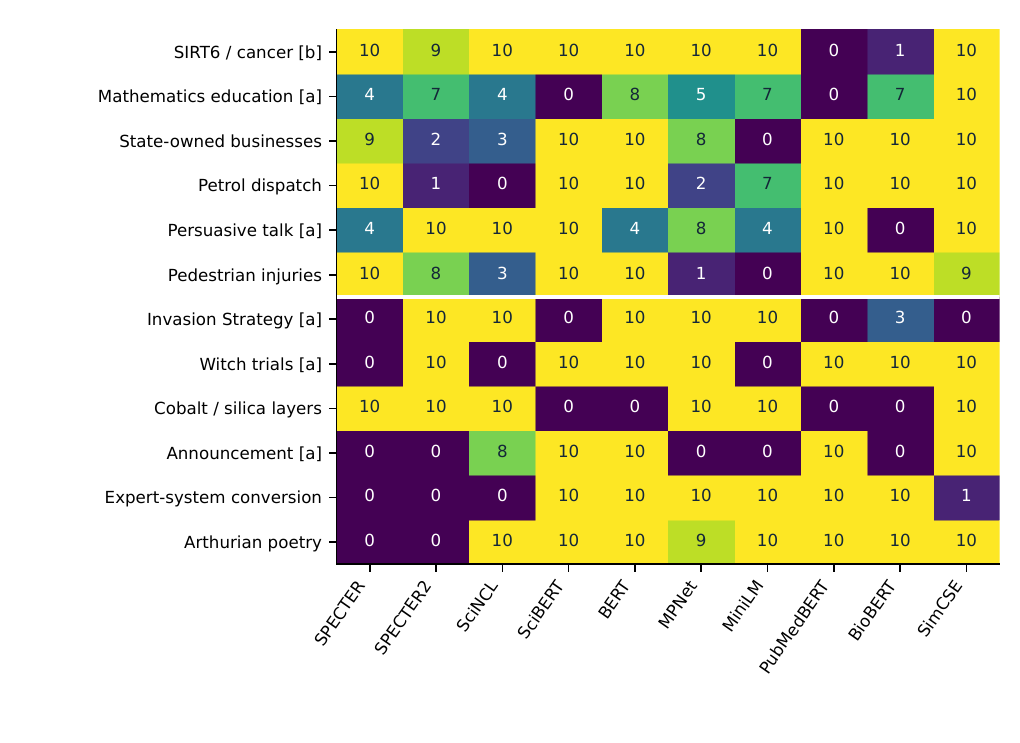}{Most encoder-dependent eligible link for each of the twelve fixed article sources. Each cell counts selections in which the target is among the 25 nearest neighbours (0--10); both endpoints are always eligible within the 256-candidate design. Upper/lower six rows are high/low agreement sources. [a] Source or label caveat; [b] the selected target has an incomplete abstract. Full IDs, titles and caveats are retained in Table S8's files. Links were selected for maximal model disagreement, so this panel does not estimate its prevalence.}{relations}{1}
\FloatBarrier

\section{Time, discipline composition and model traits}
These additional comparisons ask where agreement differs: across publication periods, areas and groups of models. Their controls assess possible explanations without identifying a single cause.

\subsection{Period contrasts and search opportunities}
Period summaries give equal weight to Fields and fixed model pairs. Endpoint differences compare 2020--24 with 2000--04; the full five-period sequence determines whether the trajectory is monotone. Native candidate pools and 2,048-candidate pools are reported separately. A later identical-query comparison separates changes in selected queries from candidate count. Publications are represented by the same present-day models across all dates; differences may reflect coverage, writing, composition, text length or model exposure rather than historical changes in scientific organisation.
\par\noindent\begin{minipage}{\linewidth}
\small
\setlength{\tabcolsep}{4pt}
\renewcommand{\arraystretch}{1.13}
\captionof{table}{Temporal endpoint changes and Field heterogeneity.}\label{tab:S09}
\begin{tabular}{@{}>{\raggedright\arraybackslash}p{6.0cm}>{\raggedright\arraybackslash}p{2.4cm}>{\raggedright\arraybackslash}p{2.15cm}>{\raggedright\arraybackslash}p{2.15cm}>{\raggedright\arraybackslash}p{2.15cm}@{}}
\toprule
\textbf{Measure / candidate design} & \textbf{Mean change} & \textbf{Positive} & \textbf{Negative} & \textbf{Non-monotone} \\
\midrule
CKA, native groups & 0.0207 & 23 & 3 & 22 \\
CKA, 2,048 per group & 0.0206 & 24 & 2 & 22 \\
Distance ranks, native & 0.0304 & 22 & 4 & 19 \\
Procrustes, native & 0.0019 & 15 & 11 & 24 \\
k25, all candidates & -0.0087 & 10 & 16 & 22 \\
k25, 2,048 / mean & 0.0091 & 20 & 6 & 20 \\
k25, 2,048 / CLS & 0.0034 & 16 & 10 & 22 \\
k25, 2,048 / SEP & 0.0138 & 21 & 5 & 20 \\
\bottomrule
\end{tabular}
\end{minipage}\par
\par\smallskip{\small End minus start: 2020--24 versus 2000--04. Each row describes 26 Fields with equal weight; neighbour changes are fractions, so multiply by 100 for percentage points. Five periods are used to classify monotonicity. Full k10/k25/k50, recipe, identical-query and pair-level checks are in the CSVs. Candidate counts can change the sign. These are descriptive period contrasts, not causal historical convergence.\par}

\PanelFigure{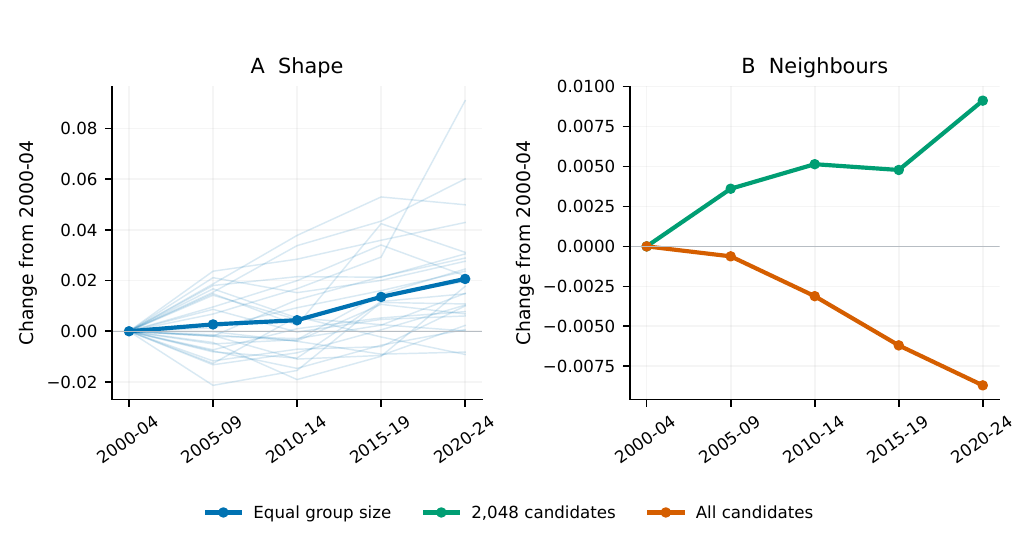}{Descriptive time patterns under candidate controls. Values are changes from 2000--04. Fields and fixed model pairs have equal weight. In (A), thin lines show the 26 Fields; the thick line is their mean. In (B), k25 overlap is compared with all candidates or 2,048 candidates. Changes in search opportunity can reverse the average sign. These are not causal historical trajectories.}{time}{1}
\subsection{Medicine and Subfield composition}
Composition controls draw 2,048 articles per Field with identical date quotas under observed or balanced Subfield composition, repeated ten times. They are also summarised after excluding BioBERT and PubMedBERT. Energy has only one eligible Subfield in this control, so its composition does not change in the same way as Medicine's. Balancing and model omission are descriptive sensitivities, not interventions isolating independent causes. The complete 26-Field results accompany the four displayed examples.
\par\noindent\begin{minipage}{\linewidth}
\small
\setlength{\tabcolsep}{4pt}
\renewcommand{\arraystretch}{1.13}
\captionof{table}{Discipline agreement with observed and balanced Subfield composition.}\label{tab:S10}
\begin{tabular}{@{}>{\raggedright\arraybackslash}p{4.35cm}>{\raggedright\arraybackslash}p{1.55cm}>{\raggedright\arraybackslash}p{2.2cm}>{\raggedright\arraybackslash}p{2.2cm}>{\raggedright\arraybackslash}p{2.2cm}>{\raggedright\arraybackslash}p{2.2cm}@{}}
\toprule
\textbf{Field} & \textbf{Models} & \textbf{CKA obs.} & \textbf{CKA bal.} & \textbf{k25 obs.} & \textbf{k25 bal.} \\
\midrule
Medicine & Ten & 0.546 & 0.543 & 29.0\% & 29.2\% \\
Medicine & Eight & 0.564 & 0.562 & 29.5\% & 29.9\% \\
Energy & Ten & 0.772 & 0.772 & 33.2\% & 33.2\% \\
Energy & Eight & 0.774 & 0.774 & 33.4\% & 33.4\% \\
Mathematics & Ten & 0.671 & 0.644 & 29.4\% & 28.8\% \\
Mathematics & Eight & 0.691 & 0.675 & 30.8\% & 30.3\% \\
Physics and Astronomy & Ten & 0.663 & 0.674 & 33.0\% & 34.1\% \\
Physics and Astronomy & Eight & 0.665 & 0.671 & 33.3\% & 34.3\% \\
\bottomrule
\end{tabular}
\end{minipage}\par
\par\smallskip{\small Eight excludes BioBERT and PubMedBERT. 2,048 articles per Field with identical date quotas; mean of ten selections. Energy has one eligible Subfield, so balancing does not introduce a comparable composition change there. The display focuses on the four Fields of Figure S5; all 26 Fields and matched contrasts are supplied. Composition and biomedical training are not isolated causal explanations.\par}

\PanelFigure{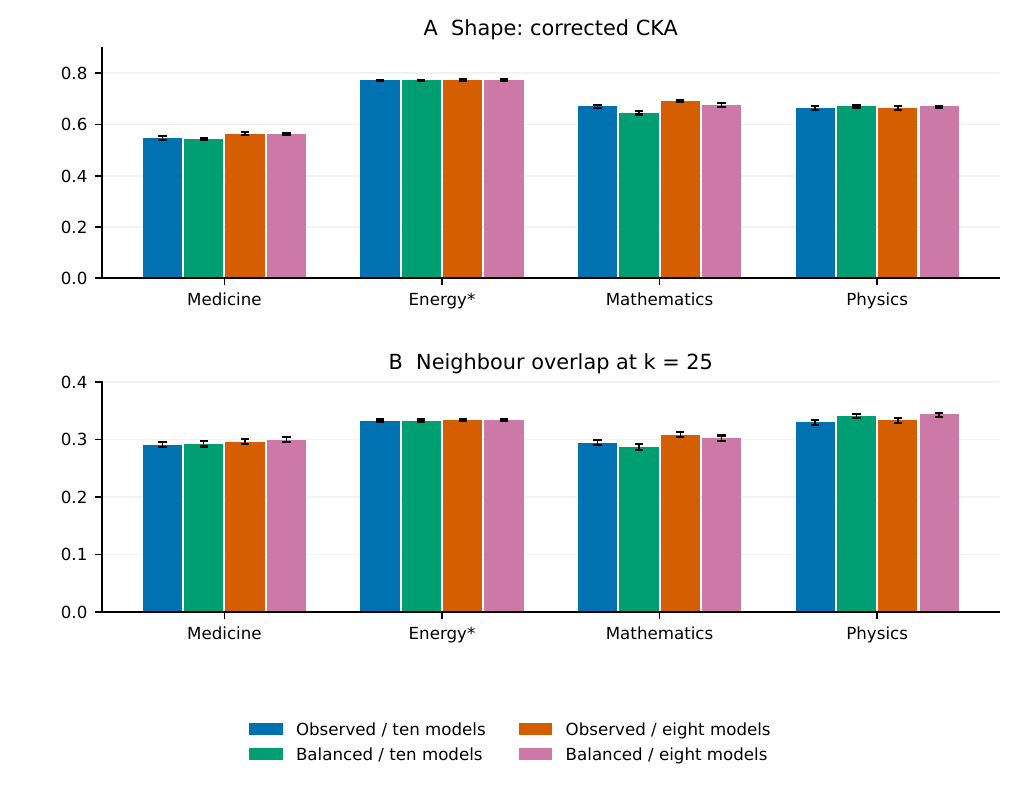}{Observed and balanced Subfield composition in four Fields. Each condition uses 2,048 articles per Field and identical date quotas. Bars are means and whiskers the range of ten selections, not confidence intervals. The eight-model panel excludes BioBERT and PubMedBERT. *Energy has one eligible Subfield. Balancing composition and omitting biomedical models do not isolate a unique causal explanation. All 26 Fields are retained in Table S10's files.}{medicine}{1}
\subsection{Readout and apparent model families}
Recorded model traits cover broad training-domain and objective groups. Pairwise models, leave-one-model checks and readout scenarios assess their descriptive association with agreement. Shared model membership makes the 45 pairs dependent. Some traits are linked in this panel, and coefficients can become non-identifiable under omission; for example, the SEP scenario without SimCSE retains the missing coefficients rather than reporting spurious estimates. The heatmaps show how apparent families depend on the representation recipe.
\par\noindent\begin{minipage}{\linewidth}
\small
\setlength{\tabcolsep}{4pt}
\renewcommand{\arraystretch}{1.13}
\captionof{table}{Descriptive model-trait fits and identifiability.}\label{tab:S11}
\begin{tabular}{@{}>{\raggedright\arraybackslash}p{5.5cm}>{\raggedright\arraybackslash}p{1.2cm}>{\raggedright\arraybackslash}p{2.3cm}>{\raggedright\arraybackslash}p{1.8cm}>{\raggedright\arraybackslash}p{4.3cm}@{}}
\toprule
\textbf{Outcome} & \textbf{Rank} & \textbf{Condition no.} & \textbf{R squared} & \textbf{Unidentified coefficients after omission} \\
\midrule
CKA / primary & 7/7 & 7.91 & 0.741 & 0 \\
CKA / 2,048 articles & 7/7 & 7.91 & 0.741 & 0 \\
CKA / quality filter & 7/7 & 7.91 & 0.752 & 0 \\
CKA / CLS & 7/7 & 18.59 & 0.459 & 14 \\
CKA / SEP & 7/7 & 8.33 & 0.258 & 7 \\
Neighbours k25 / mean & 7/7 & 7.91 & 0.759 & 0 \\
Neighbours k25 / CLS & 7/7 & 18.59 & 0.507 & 14 \\
Neighbours k25 / SEP & 7/7 & 8.33 & 0.442 & 7 \\
\bottomrule
\end{tabular}
\end{minipage}\par
\par\smallskip{\small Ten fixed models yield 45 overlapping pairs. R squared describes these pairs, not predictive validity. The final column counts unavailable coefficient rows across the ten leave-one-model checks, not independent models or tests. In the SEP design, omitting SimCSE makes coefficients non-identifiable; those values remain missing. Architecture, lineage, final corpus, objective, domain and readout are correlated traits. Full trait definitions, source URLs, coefficients, pair-label permutation references and omissions accompany this table. No causal training effect is claimed.\par}

\PanelFigure{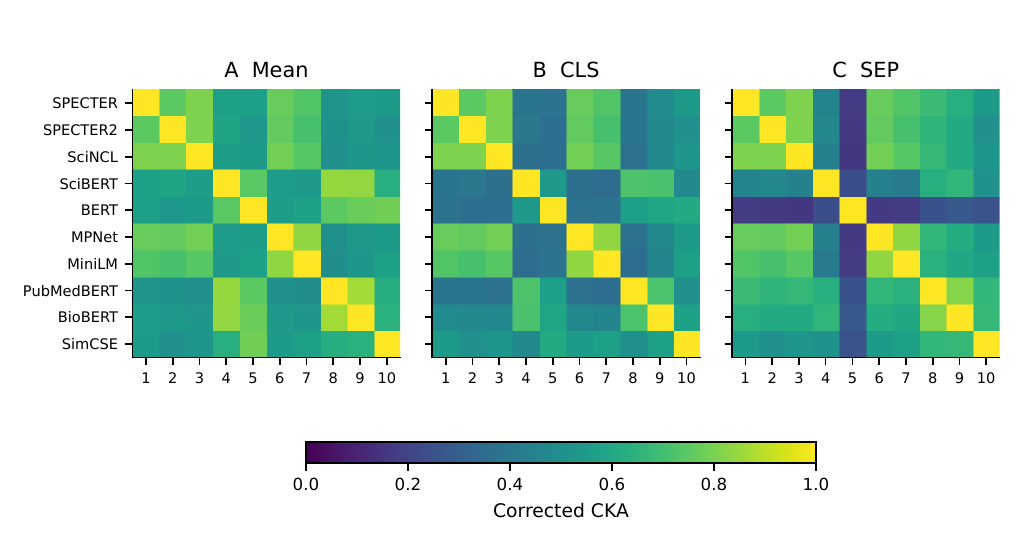}{Pairwise shape agreement under three readout choices. Matrices show mean corrected CKA over the original 130 Field-period cells. Only BERT, SciBERT, BioBERT and PubMedBERT change readout; diagonal cells are identities. Model order in columns: 1 SPECTER, 2 SPECTER2, 3 SciNCL, 4 SciBERT, 5 BERT, 6 MPNet, 7 MiniLM, 8 PubMedBERT, 9 BioBERT, 10 SimCSE. Similarity patterns are not an isolated effect of shared training.}{families}{1}
\FloatBarrier

\section{Geometric properties and construct checks}
These checks ask what a numerical property actually describes. A measure can repeat reliably and still fail to distinguish separate groups from an elongated cloud or a few extreme points.

\subsection{Properties, alternatives and selection stability}
The geometric analysis reuses the 52,000-article input panel, with 2,000 articles per Field and equal period counts. All vectors have unit length and calculations use 64-bit arithmetic. For distinct vectors $u_i,u_j$, angular spread uses the median pairwise chord distance, converted to degrees as $2\arcsin(q_{0.5}(\lVert u_i-u_j\rVert)/2)$. The saved implementation interpolates distance quantiles before this monotone conversion; the same convention is retained throughout. Alternatives are its 10th/90th percentiles and the median angle to the centre direction. They describe spread rather than semantic diversity or isotropy \citep{rudman-etal-2022-isoscore}.

After centring the unit-vector cloud, let $\lambda_j$ be the covariance eigenvalues and $p_j=\lambda_j/\sum_l\lambda_l$. The participation ratio is $\mathrm{PR}=1/\sum_j p_j^2=(\mathrm{tr}\,C)^2/\mathrm{tr}(C^2)$, where $C$ is the centred covariance matrix. This is the main measure of effective linear dimension. The entropy alternative is $\exp(-\sum_j p_j\log p_j)$, applied to the covariance spectrum \citep{roy_2007_40328}. D80 counts the smallest number of descending eigenvalues containing 80\% of variance. These quantities are unchanged by adding zero coordinates, so the 384- versus 768-coordinate comparison is not normalised by ambient dimension. They describe linear variation, not nonlinear intrinsic dimension or a count of topics.

Twenty half samples contain 1,000 articles per Field, and five further equal-size selections contain 2,000, all drawn from the fixed corpus with matched IDs across models. Central 90\% ranges and absolute median shifts are divided by the primary value. Operational thresholds are 5\% for angular spread, 10\% for PR and 20\% for connection. With five repeats, quantiles are coarse; individual values are retained. Passing those selections does not erase the four PR half-sample alerts. The complete analysis contains 16,884 sets of measures, including time, size and representation controls; not every alternative was repeated under every selection.
\par\noindent\begin{minipage}{\linewidth}
\small
\setlength{\tabcolsep}{4pt}
\renewcommand{\arraystretch}{1.13}
\captionof{table}{Selection stability of the geometric properties.}\label{tab:S12}
\begin{tabular}{@{}>{\raggedright\arraybackslash}p{3.5cm}>{\raggedright\arraybackslash}p{3.7cm}>{\raggedright\arraybackslash}p{1.2cm}>{\raggedright\arraybackslash}p{1.2cm}>{\raggedright\arraybackslash}p{2.5cm}>{\raggedright\arraybackslash}p{2.6cm}@{}}
\toprule
\textbf{Property} & \textbf{Selection design} & \textbf{Cases} & \textbf{Alerts} & \textbf{Median width} & \textbf{Maximum width} \\
\midrule
Angular spread & 20 half samples & 260 & 0 & 0.79\% & 1.81\% \\
Angular spread & 5 equal-size selections & 260 & 0 & 0.49\% & 1.85\% \\
Effective linear dimension & 20 half samples & 260 & 4 & 4.06\% & 12.84\% \\
Effective linear dimension & 5 equal-size selections & 260 & 0 & 2.60\% & 7.53\% \\
Connection & 20 half samples & 260 & 258 & 18.62\% & 423.07\% \\
Connection & 5 equal-size selections & 260 & 51 & 10.59\% & 289.69\% \\
\bottomrule
\end{tabular}
\end{minipage}\par
\par\smallskip{\small A case is one model-Field combination. Width is the central 90\% range divided by the primary value; the screen also checks median shift. Thresholds: 5\% angular spread, 10\% PR, 20\% connection. These are operational sensitivity thresholds, not journal standards or population precision. The four PR half-sample alerts remain visible even though the five equal-size selections pass. PR is effective linear dimension, not number of topics. Spectral alternatives share the same covariance spectrum and do not independently validate thematic meaning.\par}

\FloatBarrier
\subsection{Sample size}
This check varies the number of articles per area while keeping the model and measure fixed. Figure S7 shows how far each value is from the result at 2,000 articles per area.

\PanelFigure{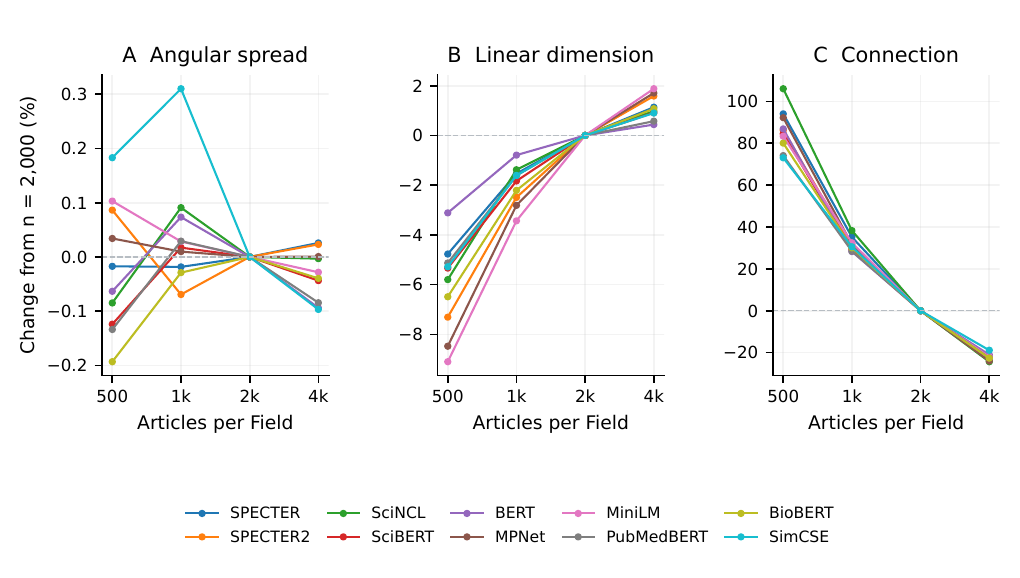}{Sample-size sensitivity of the three saved geometric properties. Each line shows one model's median relative change across the 26 Fields, compared with 2,000 articles per Field. Panels use different vertical scales. Connection uses the union-graph spectral gap with k = 25 fixed as n changes; the separate approximately fixed-fraction check is supplied in Table S13's files. These curves do not certify population precision.}{sample-size}{1}
\FloatBarrier
\subsection{Representation and content controls}
Controls vary quality flags, title/abstract input, readout, common content, extreme points, global centring and the MiniLM token limit. The quality comparison replenishes each period to its specified size after applying recorded flags. The extreme-point control removes the ten points farthest from the centre in each period of the initial half sample; removing ten random points per period supplies its equal-size reference. It does not presume that extreme points are erroneous.

Global centring subtracts the mean of the balanced 52,000 unit vectors and then normalises each result again. This changes the representation and is distinct from centring a covariance matrix for PR. Three Gaussian clouds per model--Field case provide geometric references with matching mean/covariance only in expectation before unit normalisation. Their realised moments are checked; these references are neither exact matched-moment controls nor calibrated statistical null distributions. The point controls in Figure S8 and Table S16 have their own matched references and must not be treated as another complete set of repeated selections.
\PanelFigure{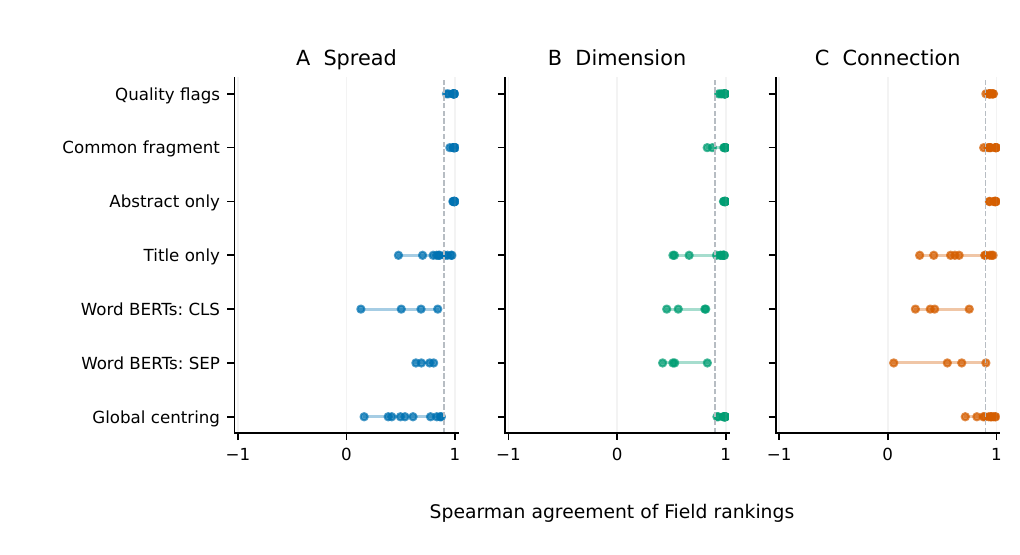}{Agreement of Field rankings before and after each matched control. Each point is a model's Spearman rank agreement; horizontal spans show their range. CLS/SEP change only the four word BERTs. The dashed line at 0.9 is a visual reference, not a pass/fail threshold. Global centring changes the representation. All saved controls, including additional point sensitivities, accompany Table S12.}{rank-controls}{1}
\FloatBarrier
\subsection{Why connection is not general fragmentation}
Connection uses the second-smallest eigenvalue of the normalised graph Laplacian $L=I-D^{-1/2}AD^{-1/2}$ on the unweighted union of directed nearest-neighbour links. The value is zero for a disconnected graph. Primary $k=25$ is checked at 10 and 50, with a weighted local-scale graph and mutual-neighbour components as alternatives. Spectral-clustering theory motivates the construction \citep{vonLuxburg2007tutorial}; it does not establish that a small gap means separate scientific themes. Eigenvalue residuals and independent dense calculations were checked.

Single-linkage merging supplies the radii at which the largest component reaches specified fractions of the cloud. Radius ratios and the maximum-to-median minimum-spanning-tree edge ratio provide different connection descriptions. Nine synthetic clouds, followed by Gaussian references and article controls, exposed non-equivalent responses to elongation, isolated outliers and separated components. Fixed $k$ also changes its fraction of candidates as sample size grows. The approximately fixed-fraction comparison was added during execution and is identified as a later sensitivity. These failures are why connection is excluded from the manuscript's interpretation of Field fragmentation.
\par\noindent\begin{minipage}{\linewidth}
\small
\setlength{\tabcolsep}{4pt}
\renewcommand{\arraystretch}{1.13}
\captionof{table}{Construct checks on all nine saved synthetic clouds.}\label{tab:S13}
\begin{tabular}{@{}>{\raggedright\arraybackslash}p{5.2cm}>{\raggedright\arraybackslash}p{2.25cm}>{\raggedright\arraybackslash}p{2.55cm}>{\raggedright\arraybackslash}p{2.7cm}>{\raggedright\arraybackslash}p{2.0cm}@{}}
\toprule
\textbf{Cloud} & \textbf{Gap k25} & \textbf{Radius 90/50} & \textbf{Max/median edge} & \textbf{Mutual giant} \\
\midrule
One round cloud & 0.499 & 1.121 & 1.371 & 96.6\% \\
One narrow cloud & 0.503 & 1.132 & 1.400 & 95.5\% \\
One wide cloud & 0.487 & 1.084 & 1.262 & 99.8\% \\
One elongated cloud & 0.091 & 1.115 & 1.299 & 99.1\% \\
One low-rank cloud & 0.098 & 1.589 & 4.913 & 99.9\% \\
Two separated groups & 0.000 & 1.766 & 2.253 & 49.5\% \\
Four separated groups & 0.000 & 1.003 & 1.587 & 24.9\% \\
Two groups with bridges & 0.007 & 1.005 & 1.428 & 97.8\% \\
One cloud with outliers & 0.499 & 1.123 & 4.770 & 95.7\% \\
\bottomrule
\end{tabular}
\end{minipage}\par
\par\smallskip{\small Each saved cloud has 1,000 points. Smaller gap can arise without separate groups; radius/edge ratios can respond to outliers or elongated clouds. In all 260 primary model-Field cases the union k25 graph is connected. Connection therefore remains a graph description and is not validated as general thematic fragmentation. The five equal-size selections retain 51 connection alerts (Table S12).\par}
\par\smallskip{\small The 780 Gaussian reference rows are included. Observed-relative reference errors span -4.69\% to 4.60\% for angular spread, and -10.47\% to 19.78\% for PR. Gaussian draws preserve fitted moments only in expectation before renormalisation; they are not calibrated null tests. Full n/k controls accompany the table; the fixed-neighbour-fraction check was added during execution.\par}

\PanelFigure{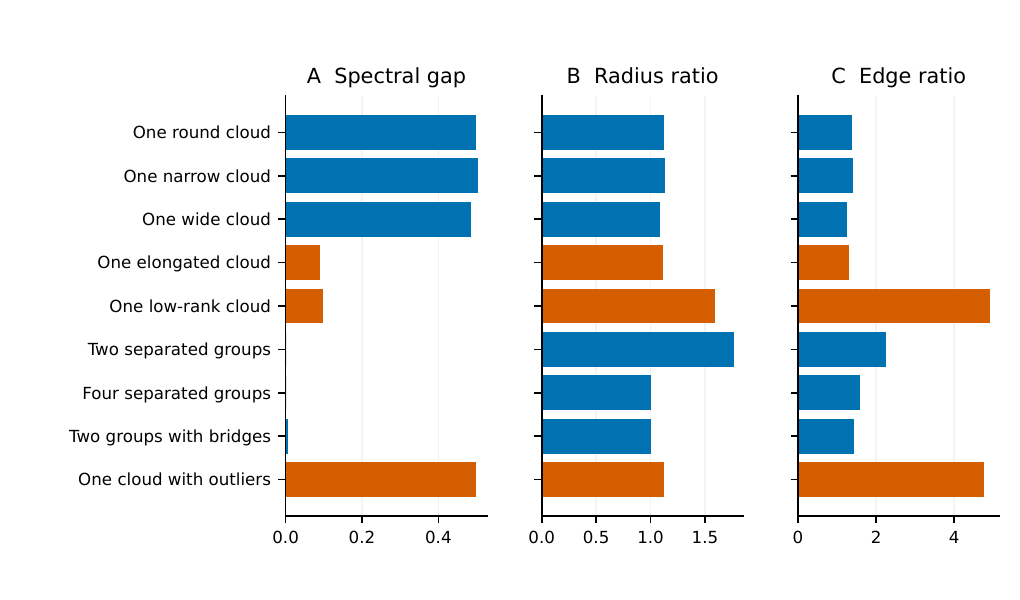}{All nine saved synthetic-cloud checks. (A) Spectral gap of the union k25 graph; smaller means weaker connection. (B) Ratio of radii at which the largest component contains 90\% and 50\% of points. (C) Maximum divided by median minimum-spanning-tree edge length. Orange marks the elongated, low-rank and outlier clouds: apparent weak connection or late merging need not mean separate thematic groups. These are controlled geometric examples, not a calibrated statistical null.}{counterexamples}{1}
\FloatBarrier

\section{All Field-pair comparisons}
Here, a pair is simply two different research areas, such as Medicine and Energy. The 325 pairs cover every such comparison among the 26 areas. The calculation asks which area has the larger geometric value, and whether changing the model reverses that answer.

\subsection{Directions and magnitude}
All unordered Field pairs are enumerated by ascending OpenAlex ID. For each model and property, the 26 saved conditions are the primary selection, twenty half samples and five additional equal-size selections, with repeat indices matched across the two Fields. If $d_r=2(B_r-A_r)/(A_r+B_r)$, direction is positive when $\min_r d_r>t+10^{-10}$ and negative when $\max_r d_r<-t-10^{-10}$. Otherwise it is unresolved. The minima $t\in\{0,0.01,0.05,0.10\}$ were specified before the pair counts, after the morphology results. They are magnitude sensitivities, not tests of significance.

All-ten agreement requires every model to have the same persistent direction. Opposition requires at least one persistently positive and one persistently negative model. All remaining pairs are unresolved, including partial agreement. These classes are exhaustive and disjoint. The full range, rather than a favourable central interval, determines persistence; the overlapping selections do not constitute independent replications.
\par\noindent\begin{minipage}{\linewidth}
\small
\setlength{\tabcolsep}{4pt}
\renewcommand{\arraystretch}{1.13}
\captionof{table}{All Field-pair conclusions under four minimum relative differences.}\label{tab:S14}
\begin{tabular}{@{}>{\raggedright\arraybackslash}p{4.3cm}>{\raggedright\arraybackslash}p{1.9cm}>{\raggedright\arraybackslash}p{2.7cm}>{\raggedright\arraybackslash}p{2.25cm}>{\raggedright\arraybackslash}p{2.3cm}>{\raggedright\arraybackslash}p{1.55cm}@{}}
\toprule
\textbf{Property} & \textbf{Minimum} & \textbf{All ten agree} & \textbf{Opposition} & \textbf{Unresolved} & \textbf{Pairs} \\
\midrule
Angular spread & 0\% & 42 & 262 & 21 & 325 \\
Angular spread & 1\% & 30 & 238 & 57 & 325 \\
Angular spread & 5\% & 10 & 96 & 219 & 325 \\
Angular spread & 10\% & 0 & 19 & 306 & 325 \\
Linear dimension & 0\% & 54 & 221 & 50 & 325 \\
Linear dimension & 1\% & 50 & 213 & 62 & 325 \\
Linear dimension & 5\% & 36 & 177 & 112 & 325 \\
Linear dimension & 10\% & 26 & 126 & 173 & 325 \\
\bottomrule
\end{tabular}
\end{minipage}\par
\par\smallskip{\small Every property uses all 325 Field pairs and ten models. A model direction must persist in all 26 saved conditions: primary sample, twenty halves and five equal-size selections. Opposition needs at least two fixed models with persistent opposite directions; unanimity needs all ten. Other cases are unresolved, not equal. The signed symmetric difference is $2(B-A)/(A+B)$; the minimum is an operational magnitude cutoff, not statistical significance or a share of science correctly represented. Full 650 pair-property rows and all 6,500 model directions accompany the table.\par}

\PanelFigure{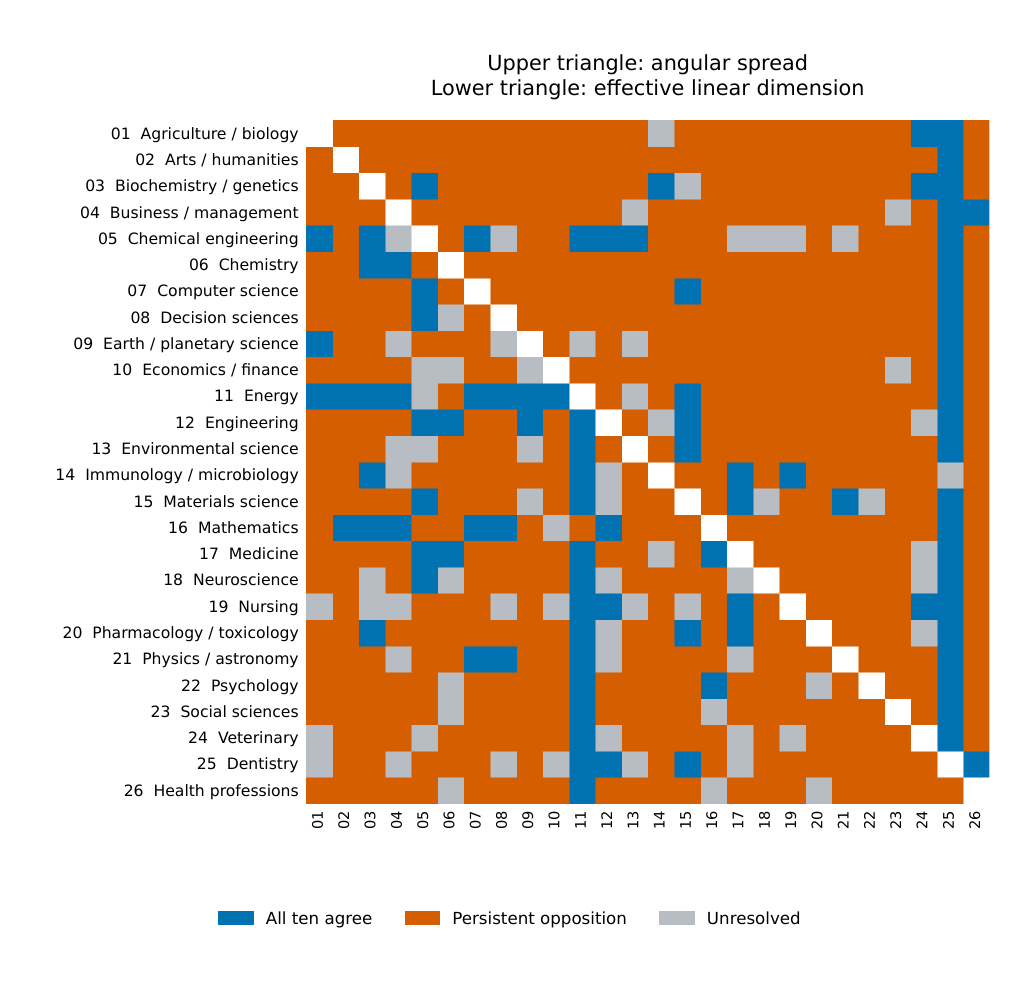}{All 650 Field-pair/property classifications at zero minimum relative difference, using the 26 saved selection conditions. The upper triangle shows angular spread; the lower shows effective linear dimension (PR). Row labels and column numbers identify the same 26 Fields; names are shortened for display. Each triangle contains 325 pairs. Blue requires all ten models to agree persistently; orange requires at least two persistently opposing models; grey is unresolved. The diagonal is not a comparison. Full labels, model directions and magnitude checks are in Table S14's files.}{all-pairs}{1}
\subsection{Alternative measures and model panels}
An opposing model pair is retained under alternatives only when those same two models preserve their original directions in every alternative. Angular alternatives, entropy dimension and D80 now all cover the primary selection, twenty half samples and five additional selections: 26 conditions each. The original three-condition dimension check is preserved as history; Table S15 uses the completed coverage. D80 is discrete and ties remain unresolved. At positive cutoffs, PR must exceed the stated minimum; its alternatives must preserve nonzero direction. Models are not replaced with different witnesses to preserve a count. Ten leave-one-model panels and the six similarity-oriented models test panel dependence; smaller panels offer fewer opportunities for opposition.
\par\noindent\begin{minipage}{\linewidth}
\small
\setlength{\tabcolsep}{3pt}
\renewcommand{\arraystretch}{1.15}
\captionof{table}{Persistent opposition with full alternative-measure coverage.}\label{tab:S15}
\begin{tabular*}{\linewidth}{@{\extracolsep{\fill}}lrrrrr@{}}
\toprule
\textbf{Model panel} & \textbf{Models} & \textbf{Spread} & \textbf{\shortstack{Spread +\\alternatives}} & \textbf{PR} & \textbf{\shortstack{PR +\\alternatives}} \\
\midrule
All ten & 10 & 262 & 225 & 221 & 190 \\[2pt]
Six similarity models & 6 & 231 & 190 & 160 & 115 \\[2pt]
Omit SPECTER & 9 & 234 & 182 & 212 & 174 \\[2pt]
Omit SPECTER2 & 9 & 251 & 205 & 211 & 173 \\[2pt]
Omit SciNCL & 9 & 259 & 217 & 215 & 182 \\[2pt]
Omit SciBERT & 9 & 258 & 222 & 218 & 187 \\[2pt]
Omit BERT & 9 & 261 & 223 & 211 & 176 \\[2pt]
Omit MPNet & 9 & 262 & 223 & 218 & 186 \\[2pt]
Omit MiniLM & 9 & 259 & 223 & 221 & 188 \\[2pt]
Omit PubMedBERT & 9 & 261 & 220 & 206 & 187 \\[2pt]
Omit BioBERT & 9 & 259 & 217 & 212 & 169 \\[2pt]
Omit SimCSE & 9 & 259 & 217 & 217 & 183 \\[2pt]
\bottomrule
\end{tabular*}
\par\smallskip{\small Counts are out of 325 at zero minimum magnitude. Every angular and dimension alternative now covers all 26 selections. The same opposing models must retain their directions; substitutes do not count. D80 ties remain unresolved. The six-model panel comprises SPECTER, SPECTER2, SciNCL, MPNet, MiniLM and SimCSE. Full cutoffs and directions are supplied as CSVs.\par}
\end{minipage}\par

\FloatBarrier
\subsection{Matched point controls}
Each of the eleven point controls is compared with its appropriate equal-size reference. Changing quality, extremes or sample size changes IDs by design; other matched controls verify identical IDs. The summary distinguishes a changed class from survival of the same original opposing models and directions in both reference and variant. One favourable controlled comparison does not demonstrate persistence across 26 new selections. The saved differences were independently audited through equivalent ratio inequalities, covering 442,000 contrasts without importing the classification function.
\par\noindent\begin{minipage}{\linewidth}
\small
\setlength{\tabcolsep}{4pt}
\renewcommand{\arraystretch}{1.13}
\captionof{table}{Point controls: retention of the original opposing models.}\label{tab:S16}
\begin{tabular}{@{}>{\raggedright\arraybackslash}p{4.3cm}>{\raggedright\arraybackslash}p{3.35cm}>{\raggedright\arraybackslash}p{2.4cm}>{\raggedright\arraybackslash}p{2.3cm}>{\raggedright\arraybackslash}p{2.3cm}@{}}
\toprule
\textbf{Control} & \textbf{Matched reference} & \textbf{Models changed} & \textbf{Spread retained} & \textbf{PR retained} \\
\midrule
Title only & half & 10 & 246/262 & 193/221 \\
Abstract only & half & 10 & 262/262 & 221/221 \\
Word BERTs: CLS & half & 4 & 253/262 & 210/221 \\
Word BERTs: SEP & half & 4 & 257/262 & 201/221 \\
Common fragment & common\_native & 10 & 262/262 & 216/221 \\
Global centring & half & 10 & 145/262 & 218/221 \\
Quality flags & half & 10 & 262/262 & 221/221 \\
Extreme-point removal & trim\_random & 10 & 262/262 & 221/221 \\
MiniLM 512 & primary & 1 & 262/262 & 221/221 \\
n = 500 & primary & 10 & 262/262 & 219/221 \\
n = 4,000 & primary & 10 & 262/262 & 221/221 \\
\bottomrule
\end{tabular}
\end{minipage}\par
\par\smallskip{\small Numerators retain the same original model pair with its original signs in both matched reference and variant. Denominators are the original persistent oppositions (262 angular spread; 221 PR). Values are not new 26-condition stability tests. The common fragment uses its own matched panel; extreme-point removal compares with random removal at equal size. Quality/extreme/size controls change IDs as specified; other matched controls retain IDs. All magnitudes and all 7,150 point pair records are included.\par}

\FloatBarrier

\section{Reproducibility and access}
\par\noindent\begin{minipage}{\linewidth}
\small
\setlength{\tabcolsep}{4pt}
\renewcommand{\arraystretch}{1.13}
\captionof{table}{Public release archives and their SHA-256 checksums.}\label{tab:S17}
\begin{tabular}{@{}>{\raggedright\arraybackslash}p{3.05cm}>{\raggedright\arraybackslash}p{10.0cm}>{\raggedright\arraybackslash}p{2.2cm}@{}}
\toprule
\textbf{Contents} & \textbf{Archive / SHA-256} & \textbf{Size} \\
\midrule
Code and guides & {\footnotesize\nolinkurl{code-and-guides.zip}}\newline{\ttfamily\seqsplit{dbc3a4b4e475866244d698abfc5f003920389b4c1ead9c6e9ed88303bbc2f104}} & 24.15 MB \\
Corpus metadata & {\footnotesize\nolinkurl{data-analysis-ready-v1.zip}}\newline{\ttfamily\seqsplit{b33ea5276587db344878cd3afc5c28c8f39274894d09e9c50e26397225a3be65}} & 44.66 MB \\
Initial comparisons & {\footnotesize\nolinkurl{data-analysis-v1.zip}}\newline{\ttfamily\seqsplit{bded3d05d03f2c2556f6d68246b99e9d5e11f9e7a614d3ef2a55263fce2f471e}} & 75.49 MB \\
Text pilot & {\footnotesize\nolinkurl{data-checklist-v1.zip}}\newline{\ttfamily\seqsplit{0615859291770cef75bdc38cf4f367c516a86bd0a7d193eca71f658cbb0739c1}} & 14.71 MB \\
Model configuration & {\footnotesize\nolinkurl{data-embeddings-v1.zip}}\newline{\ttfamily\seqsplit{cf2a01517e1e30178f4469017eadc259c36ab59d692d18e51ef92e8f5a7e6c9f}} & 0.04 MB \\
Area pairs & {\footnotesize\nolinkurl{data-field-pair-summary-v1.zip}}\newline{\ttfamily\seqsplit{71d7ddb30bed949dd01e15c8611740c10c72dcb815c8ea8309d6fd80a24afda1}} & 3.35 MB \\
Geometry & {\footnotesize\nolinkurl{data-morphology-pilot-v1.zip}}\newline{\ttfamily\seqsplit{372b07a29f77a0eb4d9ac8d34893669e7eee04efb06adc4a3cf585151d9a0836}} & 11.00 MB \\
Source cases & {\footnotesize\nolinkurl{data-prepaper-v1.zip}}\newline{\ttfamily\seqsplit{e10a813171cd3992bebd9cb5f2bd92f7e8e630a9b63d1396264ff23f05038ade}} & 2.54 MB \\
Final controls & {\footnotesize\nolinkurl{data-robustness-closure-v1.zip}}\newline{\ttfamily\seqsplit{9454a5f54f63c4dbbe16d0888778d70128d12257b34c2fb53b5c047e4c6f287b}} & 106.99 MB \\
Expanded comparisons & {\footnotesize\nolinkurl{data-robustness-v2.zip}}\newline{\ttfamily\seqsplit{7b34526a5d4097fa9e74bc39d7dc5ebebf40d313485d3c2a4e1e245b3717842e}} & 209.20 MB \\
\bottomrule
\end{tabular}
\end{minipage}\par
\par\smallskip{\small All listed archives are public under DOI \nolinkurl{10.5281/zenodo.22876602}. \nolinkurl{SELECTION.json} lists the selected data files and their origins. These are analysis-level data; historical texts and vector caches are excluded. Code is MIT, original results/documents CC BY 4.0, and OpenAlex metadata CC0.\par}

\subsection{Rebuilding this display package}
The repository command \nolinkurl{./manuscript/build.sh} compiles the included LaTeX into the main document and this supplement. It does not run models, searches or simulations. The compiler may download standard TeX support files on its first use. Re-exporting plots and table summaries is a separate presentation command, documented in the package README.

\subsection{Public statistical reproduction}
The compact data and code release is archived at \url{https://doi.org/10.5281/zenodo.22876602} \citep{treny2026reproduction}. After extracting the code and data archives together, \texttt{python }\nolinkurl{reproducibility/reproduce_analysis.py} verifies the supplied hashes and recalculates 25 tables with 106,445 rows. It checks 17,550 local neighbour scores from per-query intersection counts, recalculates the balanced local and global averages, and runs the aggregation checks for all four main figures. The regenerated table contents match the saved references exactly both locally and in a fresh Linux environment with the pinned dependencies. Pinned dependencies, commands and tolerances are provided in the release README.

This route starts from frozen model-derived measurements. It does not regenerate embeddings from historical text, reconstruct the excluded nearest-neighbour lists or certify a new run of every supplementary analysis. Historical title/abstract input tables are withheld because redistribution rights to the plaintext abstracts have not been established. Exact identifiers and input hashes are retained, but fresh OpenAlex retrieval may differ. Manuscripts and supplements are separate from the data release and may be edited without changing the archived measurements.

\clearpage
\section{Final targeted robustness checks}
These checks were specified after the previous results and the initial manuscript, before their new outcomes were inspected. They reuse the frozen vectors, preserve all earlier outputs and address the remaining selection, processing and source-quality questions. They add no model training or corpus extraction. Tables S18--S20 are accompanied by unrounded CSV summaries; the repository's \nolinkurl{data/robustness_closure_v1/} holds selections, source snapshots and per-part checksums. \nolinkurl{ROBUSTNESS_CLOSURE_PROTOCOL.md} records the design. The separate data release supports the statistical reproduction described in S8.

\subsection{Repeated centres and direct neighbour contrasts}
The 256-article centre check uses fifty selections of the same 26 Fields and 217 Subfields. Within each group, selection preserves the original five period counts. The size check fixes 26 Fields and 183 Subfields at 128, 256 and 512 articles; each has fifty repetitions. Period quotas derive from the original 512-article prefix, with nested largest-remainder rounding for smaller sizes. Thus its 256-article quotas differ from the first check. One within-period reassignment accompanies every selection, using exactly the same articles across observed/random groups and models.

The restricted comparison selects one eligible Subfield per Field fifty times, then selects articles ten times within each choice. Between-choice standard deviation of the observed mean is 0.01205; mean within-choice standard deviation is 0.00157. These are descriptive components, not independent population variances. The old restricted random reference drew 26 centres from a reassignment across all 217 Subfields; the new reference instead shuffles exactly the selected 26 groups' articles. Omitting each Field from the original centres changes mean CKA by at most 0.00687, for Mathematics; the largest individual model-pair shift is 0.02355 for MiniLM--BioBERT. All omissions are also calculated across the fifty new selections.

For direct neighbour stability, each of fifty selections draws 200 papers per Field and period from the text experiment: 1,000 candidates per Field. Queries and candidates coincide; self matches are excluded. Full exact candidate rankings can be restricted to these subsets without changing the search. Independent direct sorting checks 4,680 queries, in addition to synthetic checks with tied vectors. The contrast averages the ten same-model title changes and the 45 between-model title-plus-abstract changes, weighting Fields equally. A model-specific reference averages its nine model comparisons. Main k values are 10, 25 and 50; CLS/SEP affect only the four word BERTs and are checked at k25. Model-specific mean contrasts range from $-14.04$ to $+16.59$ percentage points; Field means range from $-4.43$ to $+5.63$. The grand mean is not a universal ordering or an equivalence test.
\par\noindent\begin{minipage}{\linewidth}
\small
\setlength{\tabcolsep}{3pt}
\renewcommand{\arraystretch}{1.15}
\captionof{table}{Repeated centres and direct input--model neighbour contrasts.}\label{tab:S18}
\textbf{A. Centres}\par\smallskip
\begin{tabular*}{\linewidth}{@{\extracolsep{\fill}}lrrrr@{}}
\toprule
\textbf{Groups} & \textbf{Papers} & \textbf{Mean} & \textbf{Selection range} & \textbf{Random mean} \\
\midrule
26 areas & 256 & 0.940 & 0.937--0.943 & 0.627 \\[2pt]
217 specialties & 256 & 0.911 & 0.910--0.912 & 0.636 \\[2pt]
26 specialties & 256 & 0.904 & 0.878--0.926 & 0.660 \\[2pt]
Areas: size check & 128 & 0.938 & 0.934--0.942 & 0.618 \\[2pt]
Areas: size check & 256 & 0.940 & 0.937--0.943 & 0.626 \\[2pt]
Areas: size check & 512 & 0.941 & 0.939--0.943 & 0.647 \\[2pt]
183 specialties & 128 & 0.913 & 0.911--0.914 & 0.617 \\[2pt]
183 specialties & 256 & 0.914 & 0.913--0.915 & 0.629 \\[2pt]
183 specialties & 512 & 0.915 & 0.914--0.915 & 0.661 \\[2pt]
\bottomrule
\end{tabular*}

\par\medskip\textbf{B. Neighbour replacement}\par\smallskip
\begin{tabular*}{\linewidth}{@{\extracolsep{\fill}}lrrrrr@{}}
\toprule
\textbf{Readout} & \textbf{k} & \textbf{\shortstack{Model\\change (\%)}} & \textbf{\shortstack{Title only\\(\%)}} & \textbf{\shortstack{Difference\\(pp)}} & \textbf{Range (pp)} \\
\midrule
Main & 10 & 70.20 & 72.17 & 1.97 & 1.85 to 2.08 \\[2pt]
Main & 25 & 64.93 & 66.49 & 1.56 & 1.44 to 1.69 \\[2pt]
Main & 50 & 60.11 & 61.15 & 1.04 & 0.94 to 1.18 \\[2pt]
CLS & 25 & 72.82 & 69.70 & -3.13 & -3.22 to -3.03 \\[2pt]
SEP & 25 & 69.15 & 67.15 & -2.00 & -2.10 to -1.88 \\[2pt]
\bottomrule
\end{tabular*}
\par\smallskip{\small Ranges are 2.5th--97.5th selection percentiles, not confidence intervals. A: 50 selections per row, except 26 specialties (50 specialty choices, ten article selections each). Size checks have their own fixed period quotas. Random groups use the same articles and dates. B: 50 selections, 1,000 candidates per area; difference = title-only replacement minus model-change replacement. No grand-mean sign reversals occur within a readout. CLS/SEP change only four word BERTs.\par}
\end{minipage}\par

\FloatBarrier
\PanelFigure{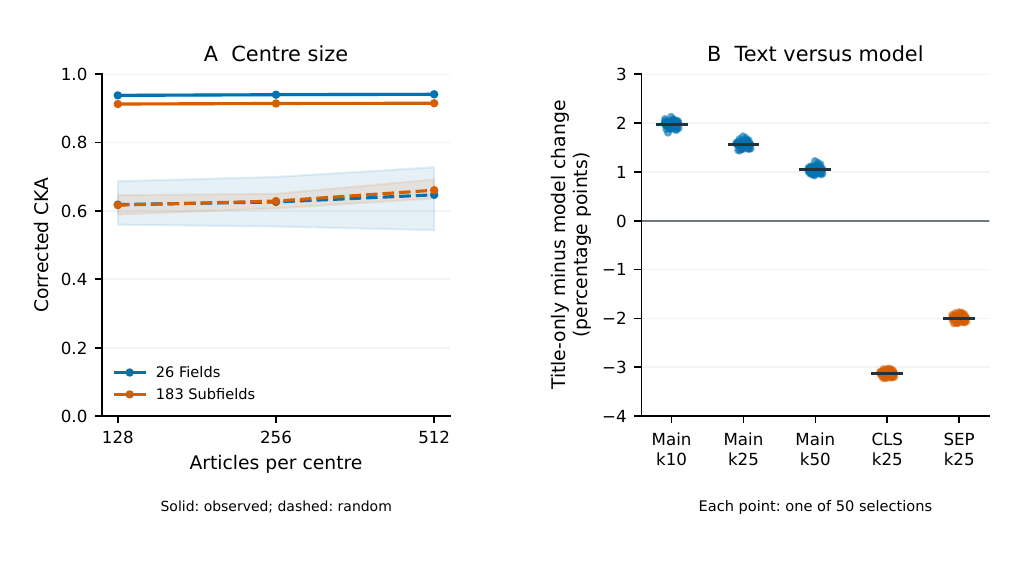}{Additional selection checks. (A) Centre agreement at three article counts; solid lines show observed groups and dashed lines show matched random groups. Shading spans 2.5th--97.5th percentiles of fifty selection means. The same 183 Subfields are used throughout. (B) Each point is one of fifty balanced selections in the direct neighbour test. Positive values mean that removing the abstract changes more neighbours than changing models; negative values mean the reverse. Black marks are medians. Main uses the primary recipes; CLS/SEP change only the four word BERTs. Neither panel supplies population confidence intervals.}{closure}{1}
\FloatBarrier

\subsection{Quality filtering with matched queries}
The existing strict filter removes seven flagged categories and retains the lowest row index in each connected component of identical nonempty DOI or normalised text. This is not a new near-duplicate detector. It retains 448,886 records. Full searches compare every surviving query against original and filtered candidates in its Field-period group. At k25, the original all-query mean is 30.29\%; restricting queries alone gives 30.58\%, and filtering their candidates gives 31.43\%. The matched-query gain is 0.85 points, or 0.77 after the existing chance adjustment.

The equal-count controls include the same 100 clean queries in both candidate pools, filling remaining places by deterministic random order. All 130 cells support 1,024 candidates; only 63 support 2,048. The largest Field-average change in full searches is 2.72 points in Mathematics. Across model pairs, the largest mean gain is 1.43 points for SPECTER2--SciBERT. The 52,000-paper text panel has at least 235 clean papers per Field-period cell. It supports 200 candidates and twenty clean queries per period in both arms, or 1,000 candidates and 100 queries per Field. Thus candidate counts, dates and queries stay matched. All k values retain substantial model disagreement. The filter's limited coverage prevents concluding that all source errors are harmless.
\par\noindent\begin{minipage}{\linewidth}
\small
\setlength{\tabcolsep}{3pt}
\renewcommand{\arraystretch}{1.15}
\captionof{table}{Quality sensitivity with the same starting papers.}\label{tab:S19}
\begin{tabular*}{\linewidth}{@{\extracolsep{\fill}}lrrrrr@{}}
\toprule
\textbf{Candidates} & \textbf{Groups} & \textbf{k} & \textbf{Original (\%)} & \textbf{Filtered (\%)} & \textbf{Change (pp)} \\
\midrule
All & 130 & 10 & 26.41 & 27.07 & 0.66 \\[2pt]
All & 130 & 25 & 30.58 & 31.43 & 0.85 \\[2pt]
All & 130 & 50 & 34.24 & 35.28 & 1.04 \\[2pt]
1024 & 130 & 10 & 30.71 & 30.91 & 0.20 \\[2pt]
1024 & 130 & 25 & 36.01 & 36.29 & 0.28 \\[2pt]
1024 & 130 & 50 & 40.80 & 41.11 & 0.31 \\[2pt]
2048 & 63 & 10 & 27.68 & 27.83 & 0.16 \\[2pt]
2048 & 63 & 25 & 32.06 & 32.25 & 0.19 \\[2pt]
2048 & 63 & 50 & 35.90 & 36.13 & 0.24 \\[2pt]
\bottomrule
\end{tabular*}

\par\medskip\textbf{Matched input check: neighbours replaced}\par\smallskip
\begin{tabular*}{\linewidth}{@{\extracolsep{\fill}}lrrrr@{}}
\toprule
\textbf{k} & \textbf{Model: original} & \textbf{Model: filtered} & \textbf{Title: original} & \textbf{Title: filtered} \\
\midrule
10 & 69.71 & 69.56 & 72.11 & 72.04 \\[2pt]
25 & 64.57 & 64.27 & 66.49 & 66.45 \\[2pt]
50 & 59.83 & 59.47 & 61.23 & 61.17 \\[2pt]
\bottomrule
\end{tabular*}
\par\smallskip{\small All values except changes are percentages. The filter retains 448,886 articles. The first block holds queries identical before and after filtering; it is distinct from the original all-query mean of 30.29\% at k25. Fixed-size searches use 100 clean queries included in both pools. Only 63 groups retain 2,048 candidates; all 130 support 1,024. The input check uses 100 queries and 1,000 date-balanced candidates per area, drawn from the saved text panel. It does not test every possible source error or near duplicate.\par}
\end{minipage}\par

\FloatBarrier

\subsection{Complete processing and dimension checks}
For each model, the global mean is calculated once from its 52,000 unit vectors. Every original unit vector in each saved selection has that mean subtracted, then is normalised again. The mean is not recalculated per Field or repetition. Covariance centring alone would leave PR unchanged; the subsequent row normalisation makes this a different representation. Spread and PR use the original main selection, twenty halves and five further equal-size selections. Entropy dimension and D80 are completed on those same 26 original selections. No selection or magnitude cutoff is changed.

The classes and fixed-model direction rules remain those of S7. At zero cutoff, 142 original angular oppositions keep the opposition class after centring, but only 128 keep an original model pair with both directions unchanged; nine new oppositions give 151 in total. PR retains 204 original oppositions and adds seven, giving 211. The same-direction requirement also retains only five of the original 54 all-ten PR agreements jointly under entropy and D80; many comparisons cannot satisfy a rule requiring every model to retain a nonzero ordering. D80 ties stay unresolved. All class transitions, reversed witness directions and magnitude checks are included, not just favourable retentions.
\par\noindent\begin{minipage}{\linewidth}
\small
\setlength{\tabcolsep}{3pt}
\renewcommand{\arraystretch}{1.15}
\captionof{table}{Full persistence after centring and under dimension alternatives.}\label{tab:S20}
\begin{tabular*}{\linewidth}{@{\extracolsep{\fill}}lrrrrrr@{}}
\toprule
\textbf{Property} & \textbf{Cutoff} & \textbf{Original} & \textbf{Centred} & \textbf{\shortstack{Same models\\and answers}} & \textbf{New} & \textbf{\shortstack{Now\\unresolved}} \\
\midrule
Spread & 0\% & 262 & 151 & 128 & 9 & 53 \\[2pt]
Spread & 1\% & 238 & 113 & 78 & 12 & 112 \\[2pt]
Spread & 5\% & 96 & 11 & 1 & 6 & 91 \\[2pt]
Spread & 10\% & 19 & 1 & 0 & 1 & 19 \\[2pt]
PR & 0\% & 221 & 211 & 204 & 7 & 16 \\[2pt]
PR & 1\% & 213 & 201 & 194 & 6 & 17 \\[2pt]
PR & 5\% & 177 & 156 & 151 & 4 & 24 \\[2pt]
PR & 10\% & 126 & 106 & 100 & 4 & 23 \\[2pt]
\bottomrule
\end{tabular*}

\par\medskip\textbf{PR oppositions retained by the alternatives}\par\smallskip
\begin{tabular*}{\linewidth}{@{\extracolsep{\fill}}lrrrr@{}}
\toprule
\textbf{PR cutoff} & \textbf{Original PR} & \textbf{Entropy} & \textbf{D80} & \textbf{Both} \\
\midrule
0\% & 221 & 209 & 190 & 190 \\[2pt]
1\% & 213 & 202 & 180 & 180 \\[2pt]
5\% & 177 & 173 & 151 & 151 \\[2pt]
10\% & 126 & 126 & 114 & 114 \\[2pt]
\bottomrule
\end{tabular*}
\par\smallskip{\small Every condition uses all 26 selections and 325 area pairs. First block: original/centred are total opposition counts; same models and answers requires an original fixed witness pair to keep both directions. New oppositions were not original oppositions. Now unresolved counts original oppositions that become unresolved. At zero cutoff, 142 angular pairs and 204 PR pairs keep the opposition class; same-witness counts are 128 and 204. Reversing both directions occurs for 19 angular witness pairs in ten area pairs, and none for PR; these counts can overlap other categories. Second block: the cutoff applies to PR, while alternatives must preserve nonzero direction in every selection. D80 ties stay unresolved. Full class transitions and witness identities are in the CSV files.\par}
\end{minipage}\par

\FloatBarrier

The numerical audit independently reproduces 78,000 model-specific directions using equivalent ratio inequalities, reproduces the saved original neighbour means to numerical precision, and verifies all 900 centre selections and matched references. None of these checks establishes thematic accuracy, independence of overlapping selections or population precision. The main text therefore distinguishes stable averages from 2,628 local selection alerts among 8,235 Subfield--model-pair comparisons.

\FloatBarrier
\bibliographystyle{plainnat}
\bibliography{references,context_references}